\documentclass[sigconf]{acmart}
\usepackage{enumitem}
\usepackage{tabularx}
\usepackage{multirow}
\usepackage{bbding}
\usepackage{pifont}
\usepackage{color}
\usepackage{subcaption}

\AtBeginDocument{%
  \providecommand\BibTeX{{%
    \normalfont B\kern-0.5em{\scshape i\kern-0.25em b}\kern-0.8em\TeX}}}

\setcopyright{acmlicensed}
\copyrightyear{2018}
\acmYear{2018}
\acmDOI{XXXXXXX.XXXXXXX}

\acmConference[Conference acronym 'XX]{Make sure to enter the correct
  conference title from your rights confirmation emai}{June 03--05,
  2018}{Woodstock, NY}
\acmISBN{978-1-4503-XXXX-X/18/06}

\begin{document}

%%
%% The "title" command has an optional parameter,
%% allowing the author to define a "short title" to be used in page headers.
\title{RA-Rec: An Efficient ID Representation Alignment Framework for LLM-based  Recommendation}

%%
%% The "author" command and its associated commands are used to define
%% the authors and their affiliations.
%% Of note is the shared affiliation of the first two authors, and the
%% "authornote" and "authornotemark" commands
%% used to denote shared contribution to the research.
\author{Xiaohan Yu}
\email{yuxh97@gmail.com}
\affiliation{%
  \institution{Huawei Poisson Lab}
  \city{Beijing}
  % \state{Ohio}
  \country{China}
  % \postcode{43017-6221}
}

\author{Li Zhang}
\email{ucesl07@ucl.ac.uk}
\affiliation{%
  \institution{University College London, \\Institute of Finance Technology}
  \city{London}
  % \state{Ohio}
  \country{United Kingdom}
  % \postcode{43017-6221}
}
\author{Xin Zhao}
\email{zhaoxin151@huawei.com}
\affiliation{%
  \institution{Huawei Poisson Lab}
  \city{Beijing}
  % \state{Ohio}
  \country{China}
  % \postcode{43017-6221}
}
\author{Yue Wang}
\email{wangyue262@huawei.com}
\affiliation{%
  \institution{Huawei Poisson Lab}
  \city{Beijing}
  % \state{Ohio}
  \country{China}
  % \postcode{43017-6221}
}
\author{Zhongrui Ma}
\email{zhongrui.ma@gmail.com}
\affiliation{%
  \institution{Huawei Poisson Lab}
  \streetaddress{P.O. Box 1212}
  \city{Beijing}
  % \state{Ohio}
  \country{China}
  % \postcode{43017-6221}
}

%%
%% By default, the full list of authors will be used in the page
%% headers. Often, this list is too long, and will overlap
%% other information printed in the page headers. This command allows
%% the author to define a more concise list
%% of authors' names for this purpose.
\renewcommand{\shortauthors}{Trovato and Tobin, et al.}

%%
%% The abstract is a short summary of the work to be presented in the
%% article.
\begin{abstract}
Large language models (LLM) have recently emerged as a powerful tool for a variety of natural language processing tasks, bringing a new surge of combining LLM with recommendation systems, termed as LLM-based RS. Current approaches generally fall into two main paradigms, the ID direct usage paradigm and the ID translation paradigm, noting their core weakness stems from lacking recommendation knowledge and uniqueness. To address this limitation, we propose a new paradigm, ID representation, which incorporates pre-trained ID embeddings into LLMs in a complementary manner. In this work, we present RA-Rec, an efficient ID representation alignment framework for LLM-based recommendation, which is compatible with multiple ID-based methods and LLM architectures. Specifically, we treat ID embeddings as soft prompts and design an innovative alignment module and an efficient tuning method with tailored data construction for alignment. Extensive experiments demonstrate RA-Rec substantially outperforms current state-of-the-art methods, achieving up to 3.0\% absolute HitRate@100 improvements while utilizing less than 10$\times$ training data.
\end{abstract}


%% A "teaser" image appears between the author and affiliation
%% information and the body of the document, and typically spans the
%% page.

% \received{20 February 2007}
% \received[revised]{12 March 2009}
% \received[accepted]{5 June 2009}

%%
%% This command processes the author and affiliation and title
%% information and builds the first part of the formatted document.

\maketitle

\section{Introduction}

Recommendation systems (RS) play an essential role in people's online lives by reducing information overload and providing users with relevant content such as news and social connections \cite{bobadilla2013recommender, sivapalan2014recommender,rahayu2022systematic}. RS learn from user-item interactions (e.g., clicks) and associated metadata (e.g., titles, user profiles) to suggest future items of interest to users \cite{resnick1997recommender, burke2007hybrid}. Recently, Large language models (LLM) have shown impressive compositional and reasoning abilities in a variety of NLP tasks, demonstrating great potential for more intelligent recommendations. As such, integrating LLMs into RS, referred to as LLM-based RS, is a valuable direction to explore. Early results provide evidence that LLM-based RS can be highly effective, particularly in cold-start and cross-domain transfer settings \cite{gao2023chat,hou2023cold,sanner2023cold}.

\begin{figure}
    \centering
    \includegraphics[width=1.0\linewidth]{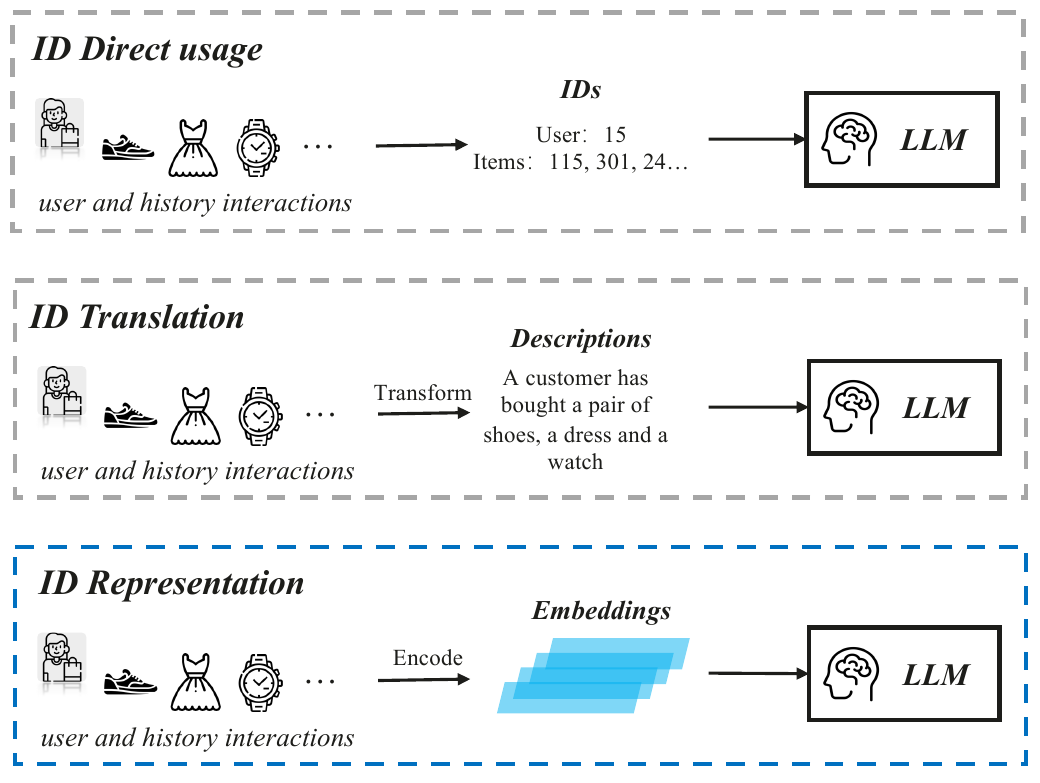}
    \caption{Three paradigms of LLM-based RS.}
    \label{fig: intro}
    % \vspace{-0.3cm}
\end{figure}

Existing attempts on LLM-based RS can be roughly classified into two paradigms: \textbf{ID Direct Usage} and \textbf{ID Translation}, as shown in Figure \ref{fig: intro}. 
In early exploration, the ID direct usage paradigm, exemplified by P5 \cite{geng2022recommendation} and its variants \cite{hua2023up5,geng2023vip5}, represents users and items as strings of ID numbers (\textit{Given the following purchase history of user 15: 115, 301, 24, predict next possible item to be purchased by the user}). However, unlike word tokens with semantics, ID numbers themselves carry no inherent meaning, rendering this paradigm vulnerable to poor generalization and transferability. Conversely, the ID translation paradigm \cite{Zhang2021,cui2022m6,wang2023generative} transforms IDs into their corresponding textual information such as titles (\textit{A customer has bought a pair of shoes, a dress and a watch, predict next possible item to be purchased by the user}) 
which is easily understood by LLMs. However, LLMs are limited to a bounded-length input (e.g., 2048 tokens for GPT3) \cite{beltagy2020longformer}, constraining their ability to take in lengthy sequences of user behaviors over time. Additionally, the natural language format in this paradigm makes it challenging to explicitly represent intricate connections between users and items such as large amount of user-item interactions in RS. 
In summary, further academic investigation is necessary to ascertain the viable methods for LLM-based RS.

User-item interaction data can be encoded into ID embeddings. These low-dimensional dense vectors retain significant collaborative information, such as complex user-item interactions and extended sequential patterns. This makes ID embeddings an important complementary to LLM in recommendation tasks, but have rarely been explored by LLM-based RS. Therefore, we propose a third paradigm, \textbf{ID Representation}, which incorporates ID embeddings into LLMs as shown in Figure \ref{fig: intro}. However, two unique challenges arise in achieving this paradigm.

\begin{enumerate}[leftmargin=*]
    \item \textit{How to bridge the gap between different modalities?} Given the heterogeneity of training tasks, ID embeddings are trained with recommendation tasks (e.g., predict the future clicked item) and LLMs are trained on NLP tasks (e.g., predict the next token), resulting in distinct modality representations. Therefore, to incorporate ID embeddings into LLMs, it is necessary to align them across modalities.
    \item \textit{How to align efficiently?} Maintaining high performance for LLM-based RS in real-world applications requires prohibitive adaptation costs in terms of both the volume of training data and the time required to train the model. This could be more challenging if LLMs need to further align different modalities as mentioned before.
\end{enumerate}

Motivated by the aforementioned challenges, we propose a novel efficient ID \textbf{R}epresentation \textbf{A}lignment framework for LLM-based \textbf{Rec}ommendation, named \textbf{RA-Rec}, that can effectively align the latent spaces of ID embeddings and LLM. RA-Rec contains three key innovations:
First, we propose a hybrid prompt approach combining soft prompts (ID embeddings providing implicit recommendation knowledge) and hard prompts (text guiding LLMs to leverage world knowledge). Second, we propose a representation alignment framework to bridge the gap between ID embeddings and LLMs. This alignment module integrates ID embeddings from various ID-based recommendation models (e.g., sequential, non-sequential) with diverse language model architectures (e.g., encoder-only, decoder-only). Third, our approach enables efficient tuning to adapt ID embeddings to LLMs.

Under the RA-Rec framework, we propose two novel alignment methods: reparameterization and contextual instruction.
To more effectively integrate recommendation system knowledge into LLMs, we propose a layer-specific projector that reparameterizes ID embeddings into soft prompts subsequently injected into each layer of the LLM.
From another perspective, ID embeddings can be viewed as contextual information for the LLM to complete sequential recommendation tasks. Rather than manually constructing prompts for each ID embedding, we propose prepending a continuous vector prefix as an instruction, guiding the LLM on how to leverage the ID information.

To enable real-world use, the proposed alignment framework should be integrated efficiently into existing LLMs with minimal computational resource consumption. We train only the alignment module rather than fine-tuning the entire LLMs, avoiding expensive retraining.
Drawing inspiration from instruction tuning techniques \cite{wei2021finetuned, zhang2023instruction,chung2022scaling}, we develop a data construction methodology that improves alignment performance by emphasizing both data quality and diversity. Specifically, we filter the data via a denoising strategy and leverage the sequence length of user behavior and item popularity as proxies for user and item diversity, respectively.
To evaluate RA-Rec, we conducted extensive experiments on real-world public datasets.
Experimental results demonstrate RA-Rec improves HitRate@100 and NDCG@10 by 25.9\% and 7.9\% on the Amazon Cloth, 21.7\% and 15.1\% on Amazon Book relatively, compared to several competitive baselines. Through visualizations and comparisons against other alignment techniques, we validated the effectiveness of the alignment module. The selective strategy also demonstrated efficiency, with under 10 times the training data and 40,000 times the parameters yielding up to a 3.0\% absolute increase in HitRate@100. The main contributions of this work are concluded as follows:

\begin{itemize}[leftmargin=*]

    \item We propose a novel paradigm that integrates ID representations learned from user-item interactions into LLMs. To our best knowledge, this work represents the first effort to incorporate ID embeddings as complements to LLM-based RS and opens avenues for LLMs to benefit from the strengths of ID-based methods while preserving user and item uniqueness.
    % new paradigm for LLM-based RS that incorporates ID embeddings learned from ID-based RS that can bring valuable RS domain knowledge and preserve the unique and comprehensive of users and items. To the best of our knowledge, it is the first attempt to treat ID embeddings as complementary to LLM-based RS.
    
    \item We develop a general alignment framework, RA-Rec, compatible with various ID-based methods and LLM architectures. We introduce an innovative alignment module as well as an efficient tuning strategy with a data construction methodology considering diversity and denoising. 
    
    \item We conduct extensive experiments on public datasets and RA-Rec consistently outperforms state-of-the-art methods, indicating the effectiveness of our method. In addition, ablation experiments and a thorough analysis of the RA-Rec are given.
\end{itemize}

\section{Related Work}
\subsection{LLM-based Recommendation}
Large language models have motivated new research on LLM-based recommendation systems. The ID Direct Usage paradigm directly incorporates IDs into LLMs. Pioneering work P5 \cite{geng2022recommendation} finetunes T5 \cite{DBLP:journals/corr/abs-1910-10683} to generate the next item ID, unifying various recommendation tasks into language generation tasks. Follow-up methods like UP5 \cite{hua2023up5} and VIP5 \cite{geng2023vip5} expand this approach by integrating additional modalities, such as images.
The ID Translation paradigm transforms IDs into textual features like titles to make them more compatible with language model inputs \cite{gao2023chat, wang2023generative, wang2023zero}. 
M6-Rec \cite{cui2022m6} representing user behavior data as plain texts and converting the recommendation tasks to either language understanding or generation. They utilize prompt tuning and other optimization techniques to minimize hardware costs.
InstructRec \cite{zhang2023instruction} poses recommendation tasks as following instructions that reflect user preferences and intentions. They defined 39 instructions and used GPT-3.5 to generate training data. 
GPT4Rec \cite{li2023gpt4rec} also adapts the natural language description of users. KAR \cite{xi2023towards} constructs hard prompts summarizing user preferences and other understandings, then uses these new features to augment existing ID models. 
% However, both paradigms are limited by sequence length constraints and struggle to fully capture nuanced ID semantics.

\subsection{Prompt Tuning for Recommendation}
Fine-tuning the Pre-trained Language Models (PLM) on recommendation specific tasks is a widely used approach that has achieved significant improvements \cite{qiu2021u, zhang2021unbert, kaviani2020emhash,wu2021empowering}. However, computational resource consumption becomes prohibitive as the size of PLMs scales up, such as Flan-T5 \cite{chung2022scaling} and LLaMA \cite{touvron2023llama} etc. 
Prompt tuning serves as an effective paradigm for adapting PLMs for recommendation tasks with the help of task-specific prompts, rather than fine-tuning all the PLM parameters. Prompts can be classified into two categories: hard and soft prompts.
Hard prompts are explicit texts that provide LLMs with information about the task at hand \cite{NEURIPS2020_1457c0d6,wei2021finetuned}. As mentioned, the majority of current LLM-based recommendation models reformulate the recommendation task as hard prompt \cite{wu2023survey}, which generates and updates the discrete text in templates during the LLM tuning with recommendation tasks \cite{dong2022survey,shin2020autoprompt}. In contrast, soft prompts are continuous vectors and can be seen as "\textit{virtual tokens}" without constraints to be word embeddings \cite{Qin2021LearningHT}. They can encode a variety of information \cite{devlin2018bert,liu2019roberta, touvron2023llama} and can be learned through back propagation \cite{li2021prefix, DBLP:journals/corr/abs-2110-07602, houlsby2019parameter, lester2021power}. While widely used in NLP, soft prompts are rarely exploited in LLM-based RS.

 \begin{figure*}[!t]
 \centering
 \includegraphics[width=15cm]{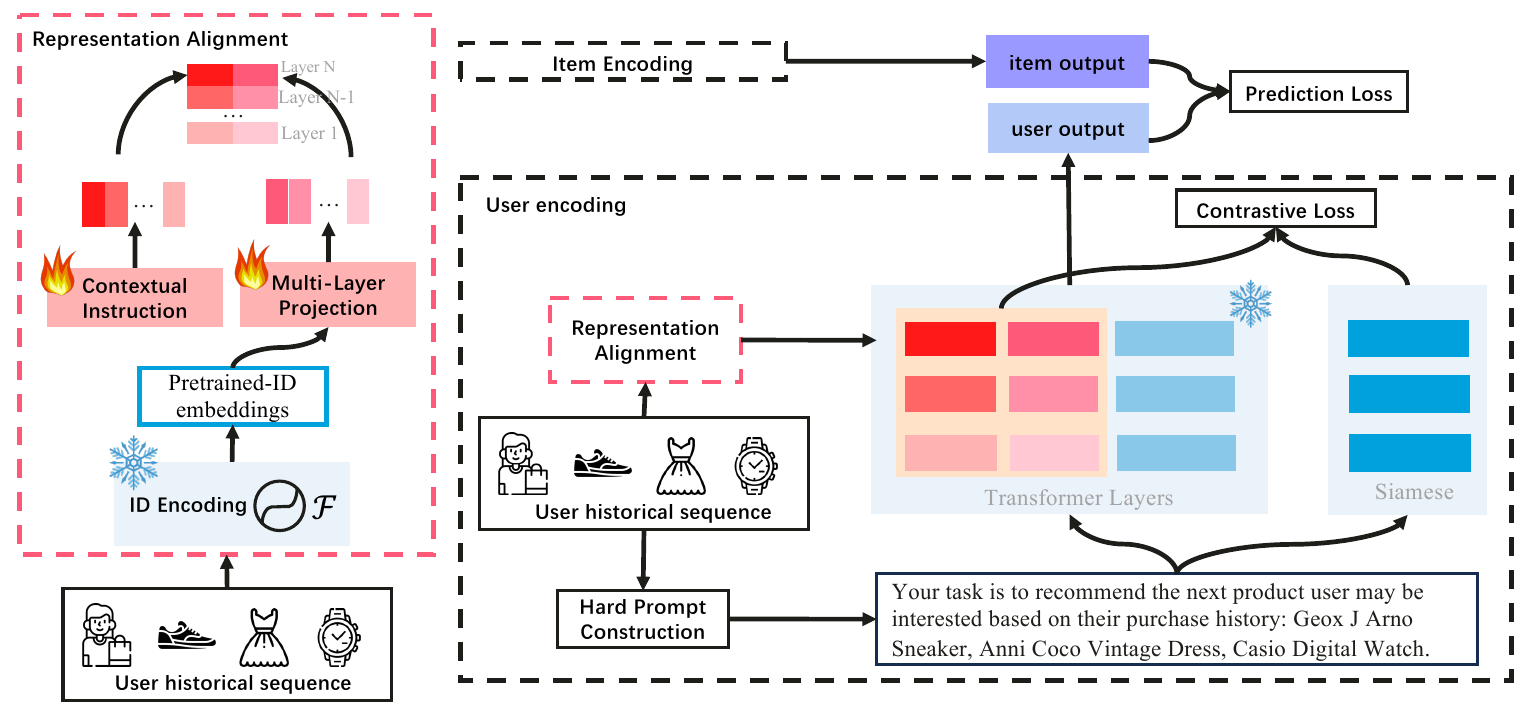}
 \caption{The overall framework of RA-Rec. } 
 % We construct a hybrid prompt through hard prompt construction module and ID Representation Alignment module detailed on the left. Then we adopt two objectives, a prediction task and a contrastive learning task during training.
 \label{fig: model architecture}
 \end{figure*}

\subsection{Multimodal Learning}
Multimodal learning aims to develop unified models that can process and integrate information from diverse modalities, including text, images and audio \cite{jabeen2023review,xu2023multimodal} etc. Early works primarily focused on multimodal pret-raining, training different models jointly to learn shared representations across data modalities \cite{sun2019videobert, li2019visualbert,,radford2021learning}. Motivated by the remarkable success of LLMs, multimodal large language models (MLLMs) have emerged as a new research frontier \cite{devlin2018bert, liu2019roberta, touvron2023llama}. Rather than pre-training MLLMs from scratch, recent works have focused on efficiently tuning LLMs while keeping their parameters frozen. A common approach in MLLM is to sum or concatenate embeddings from different modalities to create a unified representation which is then projected to match the LLM input dimensionality\cite{zeng2023matters,lyu2023macaw,huang2023language,peng2023kosmos}. Another widely used technique is applying a hierarchical or cross-attention mechanisms to model inter-modal interactions \cite{alayrac2022flamingo,lu2019vilbert,hasan2021humor}. These paradigms have proven effective and inspired various works generalizing MLLMs to other modalities such as video \cite{zhang2023video, li2023llava}, 3D-motion \cite{li2021ai} etc. However, aligning recommendation models with LLMs remains relatively underexplored in research.
Unlike visual (e.g., an image of a dog) or textual (e.g. the word "dog") information, which can be straightforwardly aligned in a shared semantic space, recommendation data presents unique challenges. It consists of large-scale user-item interactions where the recommendation results depend not only on the input but additional knowledge within the system. This makes it difficult to directly apply existing multimodal learning methods to align recommendation models and LLMs.

\section{Methodology}
In this section, we present RA-Rec, an efficient ID Representation Alignment framework, which mainly comprises three key components: 
(1) Hybrid prompt construction leveraging both soft prompts from pre-trained ID representations and crafted hard prompts;
(2) Representation alignment module which bridges the gap between ID representations and LLMs;
(3) Efficient tuning methodology which allows fast adaptation of alignment with minimal data and time.
The overall framework is shown in Figure \ref{fig: model architecture}.

\subsection{Hybrid Prompt Construction}
This section details the hybrid prompt construction approach where the ID representations learned from user-item interactions are leveraged as soft prompts, providing LLMs with implicit recommendation knowledge. The textual description of a specific recommendation task serves as the hard prompt, guiding LLMs to provide insights reaching beyond the recommendation system by tapping into its broad world knowledge.

\subsubsection{Hard Prompt}

To design a flexible and extensible hard prompt $T$ for recommendation tasks, we consider two key components: task description and recommendation information rendering raw recommendation data to human-interpretable natural languages. 
To provide adequate context for language models to perform recommendation tasks, the prompt should explicitly specifies the objective \cite{NEURIPS2020_1457c0d6}. The prompt format elaborates on the utilized recommendation information including \texttt{<USER>}, \texttt{<ITEM>} and user-item interactions \texttt{<USER-ITEM>}. Concretely, <USER> can be textual descriptions on user profiles while <ITEM> can represent features such as the item title. 
Figure \ref{fig: model architecture} showcases a constructed hard prompt for a sequential recommendation task, \textit{"Your task is to recommend the next product user may be interested based on their purchase history: Geox J Arno Sneaker, Anni Coco Vintage Dress, Casio Digital Watch"}.
Carefully engineered hard prompts can guide LLMs to incorporate relevant world knowledge to augment recommendation tasks. This allows LLM-based RS to effectively address cold-start problems, providing more intelligent recommendations for new users and items that traditional ID-based recommendation methods struggle with.

\subsubsection{Soft prompt}
The soft prompt aims to provide LLMs with knowledge of intricate user-item interactions. 
Consider a set of $N$ users $\mathcal{U}=\{u_1, u_2, \ldots u_N\}$ interacting with a set of $M$ items $\mathcal{I}=\{i_1, i_2, \ldots i_M\}$. 
The complete set of interactions, $\mathcal{D}$, can be represented by an interaction matrix, $\textbf{S}\in \mathbb{R}^{N\times M}$. However, directly injecting this high-dimensional and sparse matrix (where $N$ and $M$ can be in the billions or millions) into large language models is computationally infeasible.
We address this challenge by leveraging an ID-based recommendation model, denoted as $\mathcal{F}$, to efficiently extract low-dimensional and dense representations of both users and items. We optimize the model's parameter $\theta$ by minimizing the loss function as following:
\begin{align}
\operatorname*{minimize}_\theta \sum_{(u,i) \in \mathcal{D}} \mathcal{L}(\hat{r_{ui}}, r_{ui}), \\
\hat{r_{ui}} = \mathcal{F}(\mathbf{u}, \mathbf{i} \operatorname{|} \theta), & 
\end{align}
where $r_{ui}$ is the label (e.g., rating score, click/no-click), $\hat{r_{ui}}$ is the prediction. The model $\mathcal{F}$ can be any suitable recommendation model (e.g. sequential models). Following model training, we obtain user and item representations, $\mathbf{u} \in \mathbb{R}^{d}$ for user $u$ and $\mathbf{i} \in \mathbb{R}^{d}$ for item $i$. These representations are then used to contextualize LLMs in recommendation tasks.

Once user and item representations are learned, their input positions within the LLM require consideration. Inspired by prompt tuning methods\cite{li2021prefix, ptuning}, we propose treating these representations as virtual tokens prepended to the original LLM hard prompt at each layer. This allows the ID information to have a more significant impact on the predictions of LLMs, leading to more personalized recommendations.

% \begin{figure}
%     \centering
%     % \includegraphics[width=0.99\linewidth]{fig/hard prompt.png}
%     \includegraphics[width=0.99\linewidth]{pdfs/hard.pdf}
%     \caption{The hard prompt template. \textcolor{red}{To modify}}
%     \label{fig: hard prompt}
% \end{figure}

\subsubsection{Hybrid Prompt}
Recent studies have highlighted the limitations of LLMs in domain-specific tasks like recommendation \cite{hou2023large,chen2023knowledge,fan2023recommender} where user behavior arises from a complex interplay of factors beyond their standard training data. Furthermore, their inherent sequence length restrictions often lead to incomplete description of user-item interactions as time evolves. Conversely, soft prompts, while abundant in recommendation domain knowledge, suffer from data dependency and struggle with new users, items, or long-tail scenarios. Therefore, we argue that the complementary nature of soft prompt ($\mathbf{u}$, $\mathbf{i}$) and hard prompt $T$ can address the aforementioned limitations. The soft prompts replace the descriptive portion of $T$ with representations embodying rich RS knowledge. For new users, items, or cold-start settings where soft prompt learning falters due to insufficient data, the hard prompt can provides stability and ensures performance. Combining these advantages, we introduce our hybrid prompt.
% \textcolor{red}{ $T(\mathbf{u}, \mathbf{i}, T_{t})$}.

\subsection{Representation Alignment} \label{sec:alignment}
Bridging the divergence between LLM and ID representations is a core challenge. We propose two novel mechanisms, reparameterization and contextual instruction, to align the latent spaces of LLM and ID representations.

\subsubsection{Reparameterization}
Recent studies have demonstrated the efficacy of reparameterization techniques in integrating external knowledge into  LLMs \cite{lyu2023macaw, hu2021lora, edalati2022krona, aghajanyan2020intrinsic}. Inspired by these works, we propose a novel reparameterization module for semantically-aware fusion of user and item information from RS with the rich linguistic features within LLMs at different layers. This approach is motivated by the fundamental observation that transformer-based LLMs capture varying levels of semantic abstraction across layers \cite{jawahar-etal-2019-bert, tenney2019bert}. We argue that injecting user and item representations into LLMs should be done in a layer-specific manner to optimally leverage each layer's distinct semantic specialization. To achieve this, we introduce a layer-specific projector that transforms user and item representations into soft prompts tailored to each layer's linguistic semantics:
\begin{align}
    \mathbf{p}_{u}^{(l)} &= \mathbf{W}_{u}^{(l)} \mathbf{u} + \mathbf{b}_u^{(l)}, \\
    \mathbf{p}_{i}^{(l)} &= \mathbf{W}_{i}^{(l)} \mathbf{i} + \mathbf{b}_i^{(l)},
    \label{eq:u_p}
\end{align}
where $l\in[0, 1,..., L]$ is the layer index of LLMs and $\mathbf{W}_{u}^l,\textbf{b}_{u}^l$ $\mathbf{W}_{i}^l,\mathbf{b}_{i}^l$ are trainable parameters.

\subsubsection{Contextual Instruction}
In addition to directly applying the reparameterized ID representation as soft prompts, we exploit them as contextual information for the recommendation task. 
To effectively leverage this information, we equip the LLM with layer-prefixes, inspired by \cite{li2021prefix}. These prefixes are learnable sequences of continuous vectors that act as virtual instructions, guiding the LLMs on how to utilize the ID information at each layer of its processing.
Since our ID reparameterization utilizes layer-wise projections, we define distinct layer-prefixes, $\mathbf{c}^{(l)}_i$ and $\mathbf{c}^{(l)}_u$, for items and users, respectively. These prefixes are prepended to the corresponding reparameterized ID embeddings at each layer, producing new contextualized representations:
\begin{align}
\mathbf{d}^{(l)}_u &= [ \mathbf{c}^{(l)}_u \operatorname{||} \mathbf{p}_{u}^{(l)} ], \\
\mathbf{d}^{(l)}_i &= [ \mathbf{c}^{(l)}_i \operatorname{||} \mathbf{p}_{i}^{(l)} ],
\end{align}
where $||$ denotes the concatenation operation.
Finally, the representation of user representation is obtained by feeding the contextualized prefixes into language models:
\begin{align}
\mathbf{h}_i^{(l)} &= \operatorname{LLM}_\phi\left(\mathbf{d}^{(l)}_u,\mathbf{h}^{(l)}_{<i}\right),
\label{eq: llm}
\end{align}
where $\phi$ represents the LLM's parameters. The self-attention in each layer is calculated as:
\begin{align}
{\rm SA(\mathbf{Q}, \mathbf{K}, \mathbf{V})} &= {\rm Softmax}(\frac{\mathbf{Q}^T\mathbf{K}}{\sqrt{d'}})\mathbf{V}.
\end{align}
Crucially, the key and value matrices are modified to incorporate both the hard prompt and the contextualized ID representations:
\begin{align}
\mathbf{Q} &= \mathbf{W}_q^T \mathbf{h}_u^{(l)}, \\
\mathbf{K} &= \mathbf{W}_k^T {\rm Concat}(\mathbf{d}_u^{(l)}, \mathbf{h}_u^{(l)}), \\
\mathbf{V} &= \mathbf{W}_v^T {\rm Concat}(\mathbf{d}_u^{(l)}, \mathbf{h}_u^{(l)}),
\end{align}
where $\textbf{W}^{l}_q$, $\textbf{W}^{l}_k$, $\textbf{W}^{l}_v$ represent the transformations used to obtain the query, key and value embeddings, and $d'$ is the dimension of the hidden layer. The calculation for item is analogous to that of user. Given the final layer outputs $\mathbf{h}_u^{(L)}$ and $\mathbf{h}_i^{(L)}$ for user $u$ and item $i$, the likelihood of user $u$ interacting with item $i$ is estimated via dot-product:
\begin{equation}
    \hat{x}_{u,i}=\left( \textbf{h}_{u}^{(L)}\right)^T\textbf{h}_{i}^{(L)}.
\end{equation}

\subsection{Optimization}
To optimize our model with recommendation task, we employ the Bayesian Personalized Ranking (BPR) \cite{rendle2012bpr} pair-wise loss function. Specifically, each training sample is configured with a user $u$, a positive item $i+$ that the user has interacted with, and a negative item $i-$ that the user has not interacted with. We maximize the prediction score as follows:
\begin{align} \label{predict}
    \mathcal{L}_{p} = \sum_{(u, i) \in \mathcal{D}} \ln \sigma\left(\hat{x}_{u,i+}-\hat{x}_{u,i-}\right)-\lambda_\Theta\|\Theta\|^{2},
\end{align}
where $\hat{x}_{u,i+}$, $\hat{x}_{u,i-}$ is the likelihood of user $u$ interaction with the positive and negative items respectively, $\lambda_\Theta$ denotes a hyperparameter to determine the weight of the regularization term.
Our goal is to align the learned ID representations of users and items with the textual feature space of LLMs. 
We obtain user and item textual representations, $\Tilde{\textbf{h}}^{l}_{u}$ and $\Tilde{\textbf{h}}^{l}_{i}$, from the same underlying language model with only the hard prompt as inputs. The ID representations, $\textbf{d}^{l}_u$ and $\textbf{d}^{l}_i$, is learned from the alignment module (Eq. \ref{eq: llm}).
Inspired by the success of recent contrastive self-supervised learning \cite{info}, we propose to empower  our model by employing the InfoNCE-based contrastive learning loss with in-batch negative sampling as follows:
\begin{align} \label{align}
    \mathcal{L}_{ua} &=-\frac{1}{NL} \sum_{l=1}^L \sum_{k=1}^N \log \frac{\exp \left(\operatorname{sim}\left(\textbf{d}^{l}_{u,k}, \Tilde{\textbf{h}}^{l}_{u,k}\right) / \tau\right)}{\sum_{j=1}^N \exp \left(\operatorname{sim}\left(\textbf{d}^{l}_{u, j}, \Tilde{\textbf{h}}^{l}_{u,j}\right)) / \tau\right)}, \\
    \mathcal{L}_{ia} &=-\frac{1}{NL} \sum_{k=1}^N \sum_{l=1}^L \log \frac{\exp \left(\operatorname{sim} \left(\textbf{d}^{l}_{i, k}, \textbf{h}^{l}_{ti, k}\right)) / \tau\right)}{\sum_{j=1}^N \exp \left(\operatorname{sim}\left(\textbf{d}^{l}_{i, j}, \textbf{h}^{l}_{ti, j}\right)) / \tau\right)},
\end{align}
where N is the number of instances in a batch, sim($\cdot$, $\cdot$) is the cosine similarity, $\tau$ is a temperature coefficient.

The final training loss is a weighted sum of prediction loss (Eq. \ref{predict}) and alignment loss (Eq. \ref{align}):
\begin{equation}
    \mathcal{L} = \mathcal{L}_{p} + \lambda(\mathcal{L}_{ua} + \mathcal{L}_{ia}),
\end{equation}
where $\lambda$ is a hyperparameter to control the contribution alignment loss towards the overall objective.

\begin{table}[]
\caption{Dataset charactericstics.}
\centering
\begin{tabular}{ccccc}
\toprule[1.5pt]
Dataset & \#User  & \#Items   & \#Interaction & Sparsity\\ \midrule
Books    & 80,000  & 1,041,948 & 4,194,561     &  99.994\%     \\
Clothing   & 100,000 & 1,023,230 & 3,391,436     &  99.996\%     \\ 
\bottomrule[1.5pt]
\end{tabular}
\label{data}
\end{table}

\begingroup
\setlength{\tabcolsep}{3pt} % Default value: 6pt
\renewcommand{\arraystretch}{1} % Default value: 1
\begin{table*}[!t]
\centering
\caption{Performance comparison of different models on Amazon Books and Clothing. \textbf{Boldface} denotes the highest score and \underline{underline} indicates the best result of all baselines.
% \textbf{Enc-MLP4Rec, Enc-ComiRec, Dec-Comirec} are our model with the different ID-based models and transformer architectures where Enc is encoder-based and Dec is decoder-based. * denotes the ID-based models selected to integrate with LLM.
}
\begin{tabular}{l | l | ccc | c | c | ccc| c |c}
\toprule[1.5pt]
\multirow{2}{*}{Category}  & \multirow{2}{*}{Models} & \multicolumn{5}{c|}{Books} & \multicolumn{5}{c}{Clothing} \\ \cmidrule{3-12} 
 & \multicolumn{1}{l|}{} & HR@10  & HR@50  & HR@100 & NDCG@10 & \multicolumn{1}{l|}{MRR@10} & HR@10 & HR@50 & HR@100 & NDCG@10 & MRR@10  \\ \midrule
{\multirow{8}{*}{ID-based}} & DSSM & 0.0468 & 0.0700 & 0.0863  & 0.0371 & 0.0342 & 0.0863 & 0.0961 & 0.1021 & 0.0753 & 0.0717 \\
 & YoutubeDNN & 0.0576 & 0.0841 & 0.1012  & 0.0447 & 0.0407 & 0.0914 & 0.0964 & 0.1095 & 0.0839 & 0.0815 \\
 & Caser & 0.0446 & 0.0625 & 0.0736  & 0.0342 & 0.0309 & 0.0877 & 0.0955 & 0.0995 & 0.0748 & 0.0753 \\
 & MLP4Rec & 0.0484 & 0.0689 & 0.0846  & 0.0385 & 0.0354 & 0.0972 & 0.1077 & 0.1138 & 0.0865 & 0.0831 \\
 & GRU4Rec & 0.0362 & 0.0555 & 0.0700 & 0.0271 & 0.0243 & 0.0915 & 0.0984 & 0.1036 & 0.0856 & 0.0837 \\
 & SasRec & 0.0365 & 0.0564 & 0.0707 & 0.0284 & 0.0260 & 0.0827 & 0.0924 & 0.0977 & 0.0733 & 0.0703  \\
 & Bert4Rec & 0.0282 & 0.0528 & 0.0720 & 0.0208 & 0.0186 & 0.0135 & 0.0321 & 0.0421 & 0.0065 & 0.0043  \\
& ComiRec & \underline{0.0671} & \underline{0.0961} & \underline{0.1146} & \underline{0.0496} & \underline{0.0441} & \underline{0.1007} & \underline{0.1091} & 0.1142 & \underline{0.0919} & \underline{0.0881} \\ 
% & S3Rec & \\
\midrule

{\multirow{4}{*}{LLM-based}} & P5 & 0.0140 & 0.0167 & 0.0167 & 0.0119 & 0.0113 & 0.0286 & 0.0309 & 0.0309 & 0.0257 & 0.0248 \\
 & TwinBert & 0.0434 & 0.0683 & 0.0831 & 0.0306 & 0.0266 & 0.0845 & 0.1056 & \underline{0.1151} & 0.0530 & 0.0429 \\
 & GPT4Rec & 0.0390 & 0.0519 & 0.0559 & 0.0254 & 0.0211 & 0.0530 & 0.0650 & 0.0771 & 0.0430 & 0.0390\\ 
 & LLaMA4Rec & 0.0122 & 0.0224 & 0.0257 & 0.0072 & 0.0056 & 0.0314 & 0.0471 & 0.0539 & 0.0186 & 0.0145\\ 
\midrule

\multirow{1}{*}{Our Method} & RA-Rec &  \textbf{0.0781} & \textbf{0.1158} & \textbf{0.1395} & \textbf{0.0571} & \textbf{0.0505} & \textbf{0.1137}  &  \textbf{0.1316}  & \textbf{0.1438} & \textbf{0.0992} & \textbf{0.0946} \\ 

% & Enc-MLP4Rec  & 0.0537 & 0.0958 & 0.1243 & 0.0359 & 0.0305 & 0.1042 & 0.1209 &  0.1318 & 0.0902 & 0.0857 \\
% \multicolumn{1}{c|}{} & Enc-ComiRec  &  \textbf{0.0781} & \textbf{0.1158} & \textbf{0.1395} & \textbf{0.0571} & \textbf{0.0505} & \textbf{0.1137}  &  \textbf{0.1316}  & \textbf{0.1438} & \textbf{0.0992} & \textbf{0.0946} \\ 
% \multicolumn{1}{c|}{} & Dec-ComiRec  &  \textbf{0.0791} & \textbf{0.1114} & \textbf{0.1320} & \textbf{0.0570} & \textbf{0.0501} & \textbf{0.1229}  &  \textbf{0.1436}  & \textbf{0.1553} & \textbf{0.1034} & \textbf{0.0971} \\ 
\bottomrule[1.5pt]

\end{tabular}
\label{table:main}
\end{table*}

\subsection{Efficient Tuning} \label{sec:efficient}
Our proposed framework necessitates an additional tuning phase to align ID representations with LLMs which should be computationally efficient. To achieve this goal, we emphasize efficiency from two perspectives: data scale and model parameters.

Due to the extensive training on large datasets, LLMs and ID-based models are inherently well-trained. Studies have demonstrated that fine-tuning LLMs on specific tasks can degrade their performance \cite{touvron2023llama, NEURIPS2020_1457c0d6,zhang2023instruction}. Therefore, we opt to freeze both the language model and representations from ID-based models, focusing solely on training the semantic alignment module parameters, specifically, $\textbf{W}_{u}^{(l)},\textbf{b}_{u}^{(l)}$, $\textbf{W}_{i}^{(l)},\textbf{b}_{i}^{(l)}$ in the ID reparameterization step and $\textbf{c}^{(l)}_i$, $\textbf{c}^{(l)}_u$ in the contextual instruction step.

To construct a high-quality dataset for training the semantic alignment module, we emphasize two key dimensions: denoising and diversity \cite{tsimpoukelli2021multimodal,NEURIPS2020_1457c0d6}.
Real-world recommendation system data often suffers from noise due to biases, sparsity, and other factors \cite{zhang2019deep,wu2022graph}. 
To ensure data quality, we employ a denoising strategy that removes noisy samples. Specifically, we calculate the word overlap between target items and the user's historical interaction sequence. If the overlap is zero, indicating a lack of relevance, the sample is discarded. This filtering process eliminates irrelevant data.
In addition to denoising, diversity is crucial for capturing the breadth of user preferences and item characteristics. User diversity reflects the variance in interests and behavior, while item diversity represents the range of attributes. To ensure that the dataset encompasses a wide spectrum of user and item profiles, we use sequence length as a proxy for user diversity. Longer interaction sequences indicate greater user engagement and a broader range of interests. Similarly, we use item popularity as a proxy for item diversity. Popular items represent the most common and widely liked items, while less popular items often cater to niche interests. We divide the data into buckets based on diversity and uniformly sample from each bucket to achieve diversity across users and items.

\section{Experiment}
In this section, we present experimental details and compare our proposed methods against existing state-of-the-art models. Through extensive experiments and analysis, we aim to address the following research questions:

\begin{itemize}[leftmargin=*]
    \item RQ1: What is the overall performance of RA-Rec compared to state-of-the-art recommendation models, and does RA-Rec exhibit robust compatibility when integrated with different ID-based models and LLMs?
    \item RQ2: How well does RA-Rec align the ID representations with LLMs?
    \item RQ3: What are the quantitative impacts of the efficient tuning method for RA-Rec?
\end{itemize}

\subsection{Experimental Setup}

\subsubsection{Datasets}
To evaluate the effectiveness and generalizability of our framework, we utilize two large-scale public datasets Amazon-Books and Amazon-Clothing \cite{ni-etal-2019-justifying} on the sequential recommendation task.
In data preprocessing, we discard invalid titles that are empty or exceed 200 characters in length. Users with fewer than 3 interactions are removed, as a minimum amount of interaction data is required to represent each user. After preprocessing, we randomly sample 100,000 users from the Clothing dataset and 80,000 users from the Books dataset. Table \ref{data} summarizes key statistics of our final preprocessed datasets.
The datasets are split into train, validation, and test sets with an 80\%, 10\%, 10\% ratio. The full training sets are utilized to train ID-based recommendation models,  while the efficient tuning stage only uses less than 10\% of the training data. In the efficient tuning stage, we maintain 200,000 and 300,000 samples for the Clothing and Books datasets respectively.

\subsubsection{Evaluation Metrics}
For evaluating sequential recommendation, we apply the leave-one-out strategy in line with previous work \cite{hidasi2015session, DBLP:journals/corr/HidasiK17, cen2020controllable}. Specifically, for each user's interaction sequence, the last two items are held out as the validation and test items respectively. The remaining items in the sequence are used for training. 
We report performance using the HitRate@K where K is set to 10, 50, 100, NDCG@10 (Normalized Discounted Cumulative Gain) \cite{NDCG} and MRR@10 (Mean Reciprocal Rank) metrics.

\subsubsection{Baseline}
We compare our proposed method against two main categories of baseline models: ID-based methods and LLM-based methods. For sequential recommendation, we select the most representative methods in each category: 
(i) ID-based Methods.
DSSM \cite{dssm} and YoutubeDNN \cite{45530} utilize deep neural models to utilize user interaction histories. 
Caser \cite{caser} employs CNN in both horizontal and vertical way to model high-order Markov Chains.
GRU4Rec\cite{hidasi2015session, DBLP:journals/corr/HidasiK17} uses RNNs to model user action sequences for session-based recommendation. SasRec\cite{DBLP:journals/corr/abs-1808-09781} employs transformer-encoder architecture to extract information from user sequential behaviors. 
MLP4Rec \cite{li2022mlp4rec} develops a multi-directional fusion scheme to coherently capture sequential correlations while ComiRec\cite{cen2020controllable} and Bert4Rec \cite{Sun2019BERT4RecSR} adopt a transformer encoder block.
(ii) LLM-based Methods.
P5\cite{geng2022recommendation} formulates recommendation as generating ID strings for predicted items using finetuned FlanT5\cite{chung2022scaling}.
TwinBert\cite{lu2020twinbert} decouples the representations of query and document with BERT and a crossing layer. 
GPT4Rec \cite{li2023gpt4rec} treats recommendation as next item title prediction with GPT-2 \cite{radford2019language} while LLaMA4Rec uses freezed LLaMA-2 \cite{touvron2023llama} as backbone. 
The baselines cover the state-of-the-art in both categories.

\subsubsection{Implementation Details}
We utilize ComiRec as ID base models to obtain pre-trained ID representations and we set the number of user interests to 1. The length of the aligned prefixes for LLMs is 2 tokens. The language model backbone is SentenceBert in Table \ref{table:main}. All parameters in RA-Rec are initiated using truncated normal distribution in the range [-0.02, 0.02] and the temperature coefficient of contrastive learning is set to 0.5. We use the AdamW optimizer with an initial learning rate of $5\times 10^{-4}$ and a linear decay strategy. All the models are trained on a single NVIDIA Geforce GTX 3090 GPU.

\setlength{\tabcolsep}{4.5pt} % Default value: 6pt
\begin{table}[!t]
\centering
\caption{\small Compatibility evaluation of RA-Rec across diverse pre-trained ID base models.}
\begin{tabular}{l | c | c c | c c}
\toprule[1.5pt]
\multirow{2}{*}{ID Model} & \multirow{2}{*}{With LLM?} & \multicolumn{2}{c|}{Books} & \multicolumn{2}{c}{Clothing} \\  \cmidrule{3-6}
&  & HR@10  & N@10 & HR@10 & N@10 \\  \cmidrule{1-6}
\multirow{3}{*}{MLP4Rec} & \XSolidBrush & 0.0484 & 0.0385 & 0.0972 & 0.0865 \\ 
 & \CheckmarkBold & {0.0537} & 0.0359 & {0.1042} & {0.0902} \\ 
& Improv.(\%)  & 10.1\% & -6.7\% & 7.2\% & 4.3\% \\  
\midrule

\multirow{3}{*}{ComiRec} & \XSolidBrush & 0.0671 & 0.0496 & 0.1007 & 0.0919 \\ 
& \CheckmarkBold  & 0.0781 & {0.0571} & 0.1137 & 0.0992 \\
& Improv.(\%) & 16.4\% & 15.1\% & 12.9\% &  9.79\% \\ 
\bottomrule[1.5pt]
\end{tabular}
\label{table:generalizability}
\end{table}

\begin{figure}
    \centering
    \includegraphics[width=0.5\textwidth]{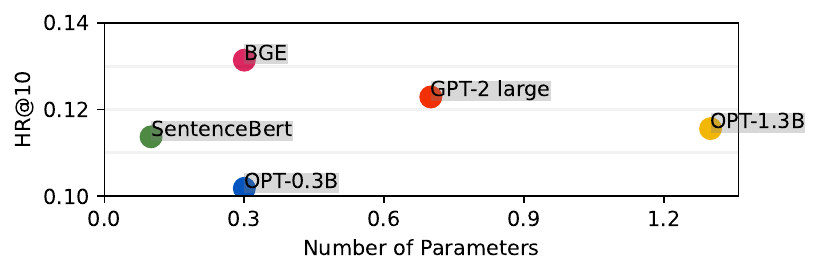}
    \caption{\small Compatibility evaluation of RA-Rec across diverse language model backbones.}
    \label{fig:compatiblity}
\end{figure}

\subsection{Performance Comparison and Compatibility (RQ1)}
In this section, we compare the overall performance of our proposed method with several state-of-the-art methods on sequential recommendation. In the meantime, our framework can be adapted for use with various ID-based models and language model architectures in a model-agnostic manner.

\subsubsection{Main Result}
The results are summarized in Table \ref{table:main}. Our method demonstrably outperforms all baseline approaches across diverse evaluation metrics, yielding statistically significant improvements for both ID-based and LLM-based methods. Notably, we achieve absolute improvements of 3.0\% in HitRate@100, 0.7\% in NDCG@10 and 0.7\% in MRR@10 compared to the best baseline. These results validate the efficacy of our proposed paradigm, which successfully incorporates ID representations into language models. This joint modeling approach effectively captures both historical interactions and textual semantics between users and items.

\subsubsection{Compatibility}
Tables \ref{table:generalizability} and Figure \ref{fig:compatiblity} showcase the compatibility experiments across different ID-based models and LLM architectures. Overall, RA-Rec exhibits remarkable compatibility with various transformer-based architectures and ID-based models. For ID-based models, RA-Rec consistently improves performance by at least 4.3\% for both MLP4Rec and ComiRec models (except for NDCG@10 in the Books dataset), highlighting the significant potential of incorporating LLMs into recommendations. We further investigated various pre-trained LLMs with diverse architectures and sizes, including SentenceBert, BGE, OPT, and GPT-2, which yields some observations: (1) The encoder-only (e.g. BGE) based LLMs may obtain relatively better results than decoder-only based LLMs (e.g. OPT-0.3B). (2) Performance improvement does not directly correlate with model size. While larger models often hold theoretical advantages, our results indicate a more nuanced relationship. This may be due to factors like scaling laws governing LLMs.

% The compatibility experiments of ID-based models and language models are shwon in Table \ref{table:generalizability} and Figure \ref{fig:compatiblity}. RA-Rec exhibits remarkable compatibility across different transformer-based architectures and ID-based models. For ID-based model, RA-Rec brings improvements of at least 4.3\% for both MLP4Rec and ComiRec models, except for NDCG@10 in Books, which indicates the great potential of introducing LLMs for recommendations. For language model backbones, we select some of the most common pre-trained langauge models with different architectures and model sizes, including SentenceBert, BGE, OPT and GPT-2. We can draw some observations: 1) The encoder-only based LLMs (e.g, BERT) obtain relatively better results than decoder-only based LLMs (e.g, GPT2); 2) While a larger parameter size theoretically suggests better performance, our results don't fully corroborate this assumption. This discrepancy could be attributed to various factors, potentially including insufficient training data or the complex scaling laws governing LLMs.

\begin{figure}[!t]
	\centering
	\begin{subfigure}{0.23\textwidth}
		\includegraphics[width=\textwidth]{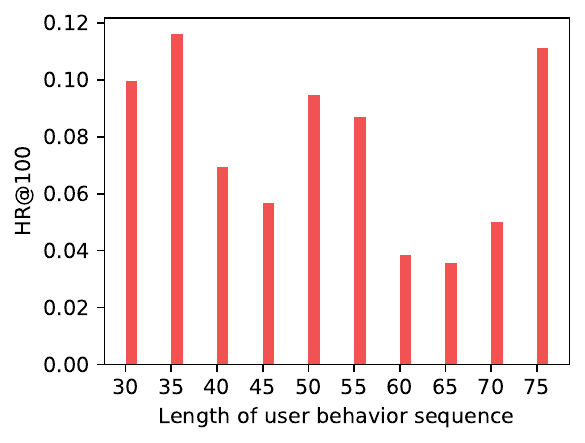}
		% \caption{\small Metrics}
		\label{fig: loss and step}
	\end{subfigure}
	\begin{subfigure}[b]{0.225\textwidth}
		\includegraphics[width=\textwidth]{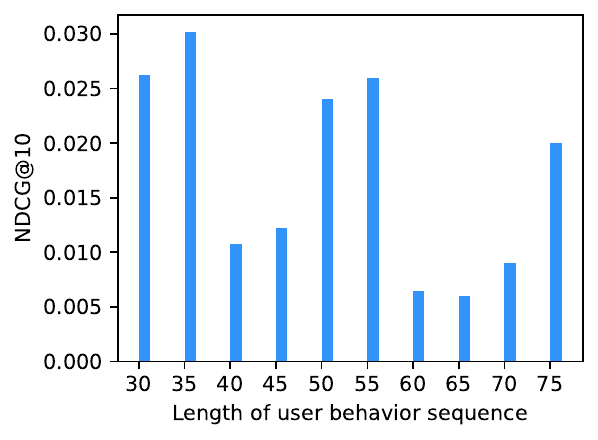}
		% \caption{\small Training Time}
		\label{fig: hit and step}
	\end{subfigure}
        \caption{\small Performance gains of RA-Rec over its base language model on data samples with long user behavior sequence lengths. The left figure shows HR@100 for accuracy and the right figure shows NDCG@10 for ranking quality. 
        }
        \label{fig:enter-label}
\end{figure}

\subsubsection{Effectiveness of ID representatiosn}
To delve deeper into the efficacy of incorporating ID embeddings within LLMs for recommendation, we compare RA-Rec (based on BERT) against BERT itself on a sequential recommendation task using the Amazon clothing dataset.
Since LLMs are limited to bound-length, but ID embeddings do not. We show model performance on longer behavior sequences, ranging from 30 to 70 in Figure \ref{fig:enter-label}. As we can see, BERT performs poorly due to input data length limitation and information loss. However, RA-Rec, with the  ID representation assistant, still performs well, proving the effectiveness of our proposed paradigm.

\begin{table}[]
\caption{ \small Performance of different alignment methods.}
\begin{tabular}{l|c c c}
\toprule[1.5pt]
Models      & HR@10                  & HR@50                  & HR@100                 \\ \midrule
RA-Rec   & \multicolumn{1}{c}{0.1137} & \multicolumn{1}{c}{0.1316} & \multicolumn{1}{c}{0.1438} \\ \midrule
RA-Rec$_{w/o\; IDRepresentation}$    & \multicolumn{1}{c}{0.0883} & \multicolumn{1}{c}{0.1172} & \multicolumn{1}{c}{0.1334} \\
RA-Rec$_{w/o\; Reparameterization}$ & \multicolumn{1}{c}{0.0857} & \multicolumn{1}{c}{0.1087} & \multicolumn{1}{c}{0.1222} \\
RA-Rec$_{w/o\; ConInstruction}$  & \multicolumn{1}{c}{0.1110} & \multicolumn{1}{c}{0.1283} & \multicolumn{1}{c}{0.1399} \\ \midrule
RA-Rec$_{Inputs}$  & \multicolumn{1}{c}{0.1046} & \multicolumn{1}{c}{0.1129} & \multicolumn{1}{c}{0.1182} \\ 
RA-Rec$_{ProjectInputs}$ & \multicolumn{1}{c}{0.0869} & \multicolumn{1}{c}{0.0965} & \multicolumn{1}{c}{0.1017} \\
\bottomrule[1.5pt]
\end{tabular}
\label{ablation:tuning}
\end{table}

\begin{figure}
    \centering
    \includegraphics[width=0.25\textwidth]{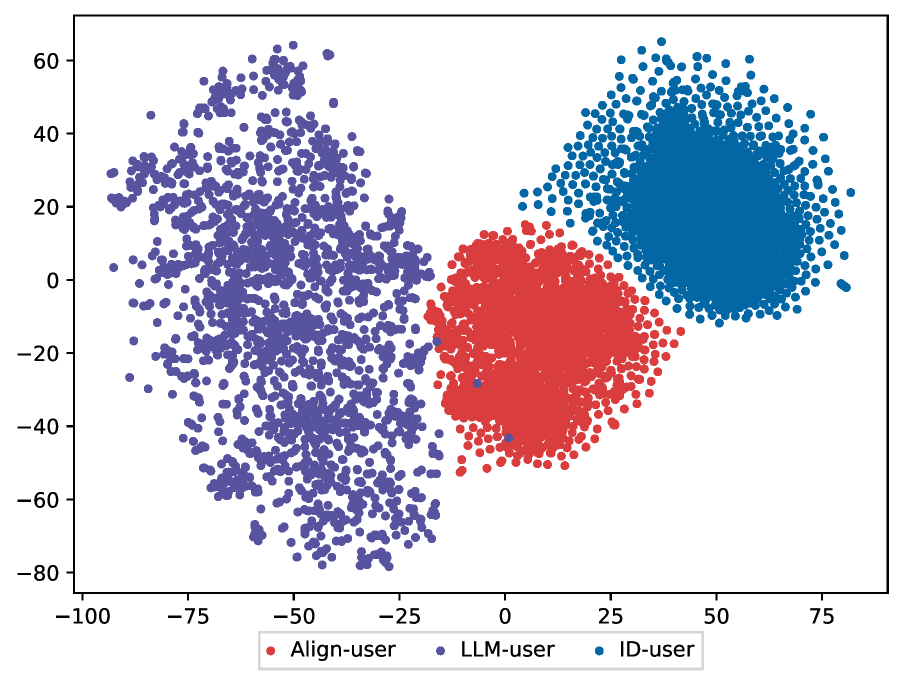}
    \caption{\small Visualization of the representations from RA-Rec, BERT and ComiRec.}
    \label{fig:visual}
    \vspace{-0.5cm}
\end{figure}

\subsection{Effectiveness of the alignment module (RQ2)}
To provide a qualitative and quantitative understanding of the alignment module in RA-Rec, we compare with several other alignment approaches and visualize the aligned representations.

\subsubsection{Comparision of numerical results}
To evaluate the effectiveness of our alignment module qualitatively, we compared it against several other alignment approaches: (i) Three variants of RA-Rec. (ii) Alignment methods proposed in other domains. To ensure fair evaluation, we kept the prefix prompt lengths consistent. Experimental results on Amazon Clothing dataset are shown in Table \ref{ablation:tuning}. 
The detailed configuration of ablated variants are as follow: RA-Rec$_{w/o\; IDRepresentation}$ excludes pre-trained ID representations which is equivalent to standard prefix tuning methods. RA-Rec$_{w/o\; Reparameterization}$ injects ID representations equally to each LLM layer. RA-Rec$_{w/o\; ConInstruction}$ removes the contextual instruction module, retaining only reparameterized ID representations. Results show that incorporating pre-trained ID representations brings significant improvement and each component of RA-Rec contributes positively to the overall performance, with the reparameterization module demonstrating more impactful improvement.
Furthermore, we identify two common alignment approaches in multimodal models: RA-Rec$_{Inputs}$ \cite{peng2023kosmos} directly feeds ID representations as inputs to LLMs and RA-Rec$_{ProjectInputs}$ \cite{lyu2023macaw} uses a self-attention module between ID representations and the LLM embedding matrix, transferring ID representations to token embeddings that LLMs can understand. Experiments demonstrate the superior performance of our alignment technique. The poorer results of these two alignment approaches indicate soft prompting is better suited than drowning the ID signals in LLMs.

% \textcolor{red}{
% Firstly, we compare against \textbf{LLM \& P-Tuning}\cite{li2021prefix} and Results show incorporating ID embeddings brings significant improvement, validating the value of multimodal fusion. Second, we conduct ablation studies on two variants, \textbf{LLM \& ConInstruct} and \textbf{LLM \& Reparam}. To ensure fair evaluation, we kept the prefix prompt lengths consistent. Comparisons reveal both alignment modules improve results, with the reparameterization module contributing more gains. 
% Finally, we identify two common alignment approaches in large multimodal models: \textbf{LLM with inputs} that directly feeds ID embeddings as inputs to LLMs and  \textbf{LLM \& Projection} \cite{lyu2023macaw} that uses a self-attention module between ID representations and the LLM embedding matrix, transferring ID representations to token embeddings that LLMs can understand. Experiments demonstrate the superior performance of our proposed technique over these two alignment methods. The poorer results of LLM with Inputs and Projection indicate soft prompting is better suited than drowning the ID signals in LLMs.
% }

\begin{table}[]
\caption{\small Effect of data size and construction methodology on RA-Rec performance.}
\setlength{\tabcolsep}{6pt} % Default value: 6pt
\begin{tabular}{l|l|ccc}
\toprule[1.5pt]
& \#Size & HR@10 & HR@50 & HR@100 \\ \midrule
\multicolumn{1}{l|}{All}                        & 330w   & 0.1104     & 0.1283     & 0.1395   \\ \midrule
\multicolumn{1}{l|}{\multirow{3}{*}{Random}}    & 20w  &  0.0851 &  0.1083  &  0.1214        \\
\multicolumn{1}{l|}{}                           & 30w  &  0.1064  &  0.1175   &  0.1253       \\
\multicolumn{1}{l|}{}                           & 50w  & 0.1111  &  0.1289  &   0.1399   \\ \midrule

\multicolumn{1}{l|}{\multirow{3}{*}{Efficient}} & 20w  &  0.1137 &  0.1316  & 0.1438  \\
 & 30w  & 0.1153 & 0.1341 & 0.1464 \\    
 & 50w  & 0.1164 & 0.1361 & 0.1487  \\
\bottomrule[1.5pt]
\end{tabular}
\label{ablation:data}
\end{table}

\subsubsection{Visualization of Alignment} \label{subsub: visualization}
To qualitatively assess the aligned representations, we projected them into a 2D space using t-SNE \cite{van2008visualizing} on Amazon Clothing dataset. Figure \ref{fig:visual} depicts the projected points, colored by model SentenceBert, ComiRec, and RA-Rec.
We observe that points for RA-Rec generally fall between the clusters formed by the other two models. This provides qualitative evidence that RA-Rec can successfully align the embedding spaces of large language models and ID representations, as intended by our alignment framework. The learned representations exhibit characteristics of both ID representations and language model, suggesting effective information fusion by RA-Rec. This visualization reinforces the ability of our framework to perform joint alignment and reasoning over multimodal inputs for recommendation systems.

\begin{table}[]
\caption{\small Training efficiency comparison of different tuning methods.}
\begin{tabular}{l| cc | cc}
\toprule[1.5pt]
\multirow{2}{*}{Methods} & \multicolumn{2}{c|}{GPT-2} & \multicolumn{2}{c}{SentenceBert} \\ \cmidrule{2-5}
 & \#Params & GPU*hrs & \#Params & GPU*hrs    \\ \midrule
% LLM$^{\textit{PT}}$         & 0.0883, 0.1172, 0.1334 & 11,392 \\ 
LLM$^{FT}$ & 774,030,080 & 32 & 109,482,240 & 7  \\
LLM$^{FT+Align}$ & 774,338,816 & 41 & 109,500,672 & 10   \\  \midrule
RA-Rec  & 18,432 & 16 & 18,432 & 3  \\
\bottomrule[1.5pt]
\end{tabular}
\label{ablation}
\end{table}

\begin{figure}[!t]
	\centering
	\begin{subfigure}[b]{0.23\textwidth}
		\includegraphics[width=\textwidth]{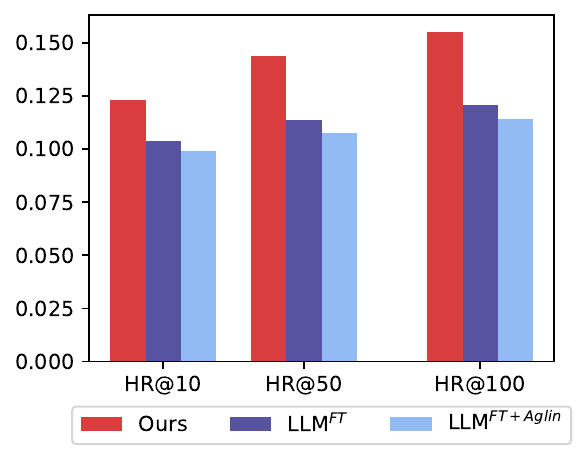}
		\caption{\small Metrics}
		\label{fig: loss and step}
	\end{subfigure}
	\begin{subfigure}[b]{0.23\textwidth}
		\includegraphics[width=\textwidth]{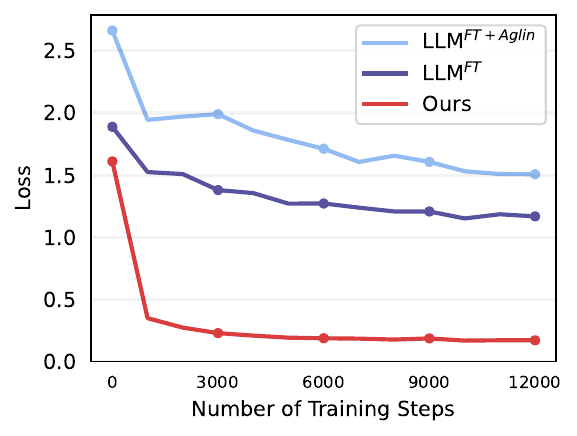}
		\caption{\small Training Loss Curve}
		\label{fig: hit and step}
	\end{subfigure}
	\caption{ \small Time efficiency evaluation on Amazon Cloth with SentenceBert as base model. The left figure shows the performance of HR@K (K=10,50,100). The right figure shows the training loss curve.}
    \vspace{-0.5cm}
	\label{fig: efficiency figure}
\end{figure}

\subsection{Effectiveness of efficient tuning (RQ3)}
To validate the effectiveness of our proposed efficient tuning method, we conduct experiments from two perspectives: data efficiency and training efficiency. 

\subsubsection{Data efficiency Study}
We propose a novel data construction method tailored for ID-LLM alignment, that considers both diversity and denoising. We conduct experiments on Amazon Clothing dataset. We aim to answer two key questions:
(i) Does our proposed data construction method improve data efficiency? 
To validate this, we train the alignment module on three datasets: our constructed \textbf{Efficient} dataset, a \textbf{Random} sampled dataset, and the full \textbf{All} dataset. As shown in Table \ref{ablation:data}, \textbf{Efficient} outperforms \textbf{Random} by 2.2\% in HR@100 with the same data scale and achieves almost 1\% higher HR@100 compared to using all data. This confirms that our our considerations of diversity and denoising effectively improves data quality, leading to better alignment modeling.
(ii) Is the constructed dataset size sufficient to train a high-quality alignment module? 
We systematically vary data size from 200,000 to 500,000 with both random and our constructed methods, training the RA-Rec model on each. Results show that using only 6\% of selected user-item interactions achieves better alignment accuracy than the entire dataset, indicating our constructed data scale sufficiently meets volume needs. RA-Rec performance continues to improve with increasing data size, reaching a HitRate@100 of 0.1487 from 0.1438. This demonstrates that a small yet carefully designed dataset is imperative to train an effective alignment module.

\subsubsection{Time Efficiency Study}
To illustrate the efficiency of our model which only tunes the alignment module, we compare with LLM$^{FT}$ which has no alignment module and finetunes LLMs while keeping ID representations frozen and LLM$^{FT+Align}$ which jointly trains LLM and the alignment module. Some statistics are shown in Table \ref{ablation} while the performance and training curve are presented in Figure \ref{fig: efficiency figure}. Our proposed approach converges significantly faster while achieving superior performance. Specifically, our model outperforms others by 1.3\%, 2.3\%, and 3.0\% on HR@10, 50 and 100, using only $2\times10^{-5}\%$ of the parameters and 50\% of the training computation compared to alternatives. During training, RA-Rec reaches near peak performance at just 7,000 steps, exhibiting a steeper decline in training loss. In conclusion, by solely tuning the lightweight alignment module, our model attains state-of-the-art results with substantially higher efficiency regarding parameters, computation, and convergence speed.

\section{Conclusion}

In this paper, we proposed RA-Rec, a novel LLM-based RS paradigm that integrates ID representations learned from user-item interactions into LLMs for improved recommendation accuracy. RA-Rec combines a hybrid prompt design, a robust representation alignment module with contrastive learning, and an efficient tuning strategy with tailored data construction. Extensive experiments on public datasets demonstrate that RA-Rec consistently outperforms state-of-the-art methods, highlighting its effectiveness in recommendation.

\bibliographystyle{ACM-Reference-Format}
\bibliography{sample-base}

%%% -*-BibTeX-*-
%%% Do NOT edit. File created by BibTeX with style
%%% ACM-Reference-Format-Journals [18-Jan-2012].

\begin{thebibliography}{81}

%%% ====================================================================
%%% NOTE TO THE USER: you can override these defaults by providing
%%% customized versions of any of these macros before the \bibliography
%%% command.  Each of them MUST provide its own final punctuation,
%%% except for \shownote{}, \showDOI{}, and \showURL{}.  The latter two
%%% do not use final punctuation, in order to avoid confusing it with
%%% the Web address.
%%%
%%% To suppress output of a particular field, define its macro to expand
%%% to an empty string, or better, \unskip, like this:
%%%
%%% \newcommand{\showDOI}[1]{\unskip}   % LaTeX syntax
%%%
%%% \def \showDOI #1{\unskip}           % plain TeX syntax
%%%
%%% ====================================================================

\ifx \showCODEN    \undefined \def \showCODEN     #1{\unskip}     \fi
\ifx \showDOI      \undefined \def \showDOI       #1{#1}\fi
\ifx \showISBNx    \undefined \def \showISBNx     #1{\unskip}     \fi
\ifx \showISBNxiii \undefined \def \showISBNxiii  #1{\unskip}     \fi
\ifx \showISSN     \undefined \def \showISSN      #1{\unskip}     \fi
\ifx \showLCCN     \undefined \def \showLCCN      #1{\unskip}     \fi
\ifx \shownote     \undefined \def \shownote      #1{#1}          \fi
\ifx \showarticletitle \undefined \def \showarticletitle #1{#1}   \fi
\ifx \showURL      \undefined \def \showURL       {\relax}        \fi
% The following commands are used for tagged output and should be
% invisible to TeX
\providecommand\bibfield[2]{#2}
\providecommand\bibinfo[2]{#2}
\providecommand\natexlab[1]{#1}
\providecommand\showeprint[2][]{arXiv:#2}

\bibitem[Aghajanyan et~al\mbox{.}(2020)]%
        {aghajanyan2020intrinsic}
\bibfield{author}{\bibinfo{person}{Armen Aghajanyan}, \bibinfo{person}{Luke Zettlemoyer}, {and} \bibinfo{person}{Sonal Gupta}.} \bibinfo{year}{2020}\natexlab{}.
\newblock \showarticletitle{Intrinsic dimensionality explains the effectiveness of language model fine-tuning}.
\newblock \bibinfo{journal}{\emph{arXiv preprint arXiv:2012.13255}} (\bibinfo{year}{2020}).
\newblock


\bibitem[Alayrac et~al\mbox{.}(2022)]%
        {alayrac2022flamingo}
\bibfield{author}{\bibinfo{person}{Jean-Baptiste Alayrac}, \bibinfo{person}{Jeff Donahue}, \bibinfo{person}{Pauline Luc}, \bibinfo{person}{Antoine Miech}, \bibinfo{person}{Iain Barr}, \bibinfo{person}{Yana Hasson}, \bibinfo{person}{Karel Lenc}, \bibinfo{person}{Arthur Mensch}, \bibinfo{person}{Katherine Millican}, \bibinfo{person}{Malcolm Reynolds}, {et~al\mbox{.}}} \bibinfo{year}{2022}\natexlab{}.
\newblock \showarticletitle{Flamingo: a visual language model for few-shot learning}.
\newblock \bibinfo{journal}{\emph{NeurIPS}}  \bibinfo{volume}{35} (\bibinfo{year}{2022}), \bibinfo{pages}{23716--23736}.
\newblock


\bibitem[Beltagy et~al\mbox{.}(2020)]%
        {beltagy2020longformer}
\bibfield{author}{\bibinfo{person}{Iz Beltagy}, \bibinfo{person}{Matthew~E Peters}, {and} \bibinfo{person}{Arman Cohan}.} \bibinfo{year}{2020}\natexlab{}.
\newblock \showarticletitle{Longformer: The long-document transformer}.
\newblock \bibinfo{journal}{\emph{arXiv preprint arXiv:2004.05150}} (\bibinfo{year}{2020}).
\newblock


\bibitem[Bobadilla et~al\mbox{.}(2013)]%
        {bobadilla2013recommender}
\bibfield{author}{\bibinfo{person}{Jes{\'u}s Bobadilla}, \bibinfo{person}{Fernando Ortega}, \bibinfo{person}{Antonio Hernando}, {and} \bibinfo{person}{Abraham Guti{\'e}rrez}.} \bibinfo{year}{2013}\natexlab{}.
\newblock \showarticletitle{Recommender systems survey}.
\newblock \bibinfo{journal}{\emph{Knowledge-based systems}} (\bibinfo{year}{2013}).
\newblock


\bibitem[Brown et~al\mbox{.}(2020)]%
        {NEURIPS2020_1457c0d6}
\bibfield{author}{\bibinfo{person}{Tom Brown}, \bibinfo{person}{Benjamin Mann}, \bibinfo{person}{Nick Ryder}, \bibinfo{person}{Melanie Subbiah}, \bibinfo{person}{Jared~D Kaplan}, \bibinfo{person}{Prafulla Dhariwal}, \bibinfo{person}{Arvind Neelakantan}, \bibinfo{person}{Pranav Shyam}, \bibinfo{person}{Girish Sastry}, \bibinfo{person}{Amanda Askell}, \bibinfo{person}{Sandhini Agarwal}, \bibinfo{person}{Ariel Herbert-Voss}, \bibinfo{person}{Gretchen Krueger}, \bibinfo{person}{Tom Henighan}, \bibinfo{person}{Rewon Child}, \bibinfo{person}{Aditya Ramesh}, \bibinfo{person}{Daniel Ziegler}, \bibinfo{person}{Jeffrey Wu}, \bibinfo{person}{Clemens Winter}, \bibinfo{person}{Chris Hesse}, \bibinfo{person}{Mark Chen}, \bibinfo{person}{Eric Sigler}, \bibinfo{person}{Mateusz Litwin}, \bibinfo{person}{Scott Gray}, \bibinfo{person}{Benjamin Chess}, \bibinfo{person}{Jack Clark}, \bibinfo{person}{Christopher Berner}, \bibinfo{person}{Sam McCandlish}, \bibinfo{person}{Alec Radford}, \bibinfo{person}{Ilya Sutskever}, {and}
  \bibinfo{person}{Dario Amodei}.} \bibinfo{year}{2020}\natexlab{}.
\newblock \showarticletitle{Language Models are Few-Shot Learners}. In \bibinfo{booktitle}{\emph{NeurIPS}}, \bibfield{editor}{\bibinfo{person}{H.~Larochelle}, \bibinfo{person}{M.~Ranzato}, \bibinfo{person}{R.~Hadsell}, \bibinfo{person}{M.F. Balcan}, {and} \bibinfo{person}{H.~Lin}} (Eds.).
\newblock


\bibitem[Burke(2007)]%
        {burke2007hybrid}
\bibfield{author}{\bibinfo{person}{Robin Burke}.} \bibinfo{year}{2007}\natexlab{}.
\newblock \showarticletitle{Hybrid web recommender systems}.
\newblock \bibinfo{journal}{\emph{The adaptive web: methods and strategies of web personalization}} (\bibinfo{year}{2007}), \bibinfo{pages}{377--408}.
\newblock


\bibitem[Cen et~al\mbox{.}(2020)]%
        {cen2020controllable}
\bibfield{author}{\bibinfo{person}{Yukuo Cen}, \bibinfo{person}{Jianwei Zhang}, \bibinfo{person}{Xu Zou}, \bibinfo{person}{Chang Zhou}, \bibinfo{person}{Hongxia Yang}, {and} \bibinfo{person}{Jie Tang}.} \bibinfo{year}{2020}\natexlab{}.
\newblock \showarticletitle{Controllable multi-interest framework for recommendation}. In \bibinfo{booktitle}{\emph{SIGKDD}}.
\newblock


\bibitem[Chen et~al\mbox{.}(2023)]%
        {chen2023knowledge}
\bibfield{author}{\bibinfo{person}{Jiao Chen}, \bibinfo{person}{Luyi Ma}, \bibinfo{person}{Xiaohan Li}, \bibinfo{person}{Nikhil Thakurdesai}, \bibinfo{person}{Jianpeng Xu}, \bibinfo{person}{Jason~HD Cho}, \bibinfo{person}{Kaushiki Nag}, \bibinfo{person}{Evren Korpeoglu}, \bibinfo{person}{Sushant Kumar}, {and} \bibinfo{person}{Kannan Achan}.} \bibinfo{year}{2023}\natexlab{}.
\newblock \showarticletitle{Knowledge Graph Completion Models are Few-shot Learners: An Empirical Study of Relation Labeling in E-commerce with LLMs}.
\newblock \bibinfo{journal}{\emph{arXiv preprint arXiv:2305.09858}} (\bibinfo{year}{2023}).
\newblock


\bibitem[Chung et~al\mbox{.}(2022)]%
        {chung2022scaling}
\bibfield{author}{\bibinfo{person}{Hyung~Won Chung}, \bibinfo{person}{Le Hou}, \bibinfo{person}{Shayne Longpre}, \bibinfo{person}{Barret Zoph}, \bibinfo{person}{Yi Tay}, \bibinfo{person}{William Fedus}, \bibinfo{person}{Eric Li}, \bibinfo{person}{Xuezhi Wang}, \bibinfo{person}{Mostafa Dehghani}, \bibinfo{person}{Siddhartha Brahma}, {et~al\mbox{.}}} \bibinfo{year}{2022}\natexlab{}.
\newblock \showarticletitle{Scaling instruction-finetuned language models}.
\newblock \bibinfo{journal}{\emph{arXiv preprint arXiv:2210.11416}} (\bibinfo{year}{2022}).
\newblock


\bibitem[Covington et~al\mbox{.}(2016)]%
        {45530}
\bibfield{author}{\bibinfo{person}{Paul Covington}, \bibinfo{person}{Jay Adams}, {and} \bibinfo{person}{Emre Sargin}.} \bibinfo{year}{2016}\natexlab{}.
\newblock \showarticletitle{Deep Neural Networks for YouTube Recommendations}. In \bibinfo{booktitle}{\emph{Proceedings of the 10th ACM Conference on Recommender Systems}}. \bibinfo{address}{New York, NY, USA}.
\newblock


\bibitem[Cui et~al\mbox{.}(2022)]%
        {cui2022m6}
\bibfield{author}{\bibinfo{person}{Zeyu Cui}, \bibinfo{person}{Jianxin Ma}, \bibinfo{person}{Chang Zhou}, \bibinfo{person}{Jingren Zhou}, {and} \bibinfo{person}{Hongxia Yang}.} \bibinfo{year}{2022}\natexlab{}.
\newblock \showarticletitle{M6-rec: Generative pretrained language models are open-ended recommender systems}.
\newblock \bibinfo{journal}{\emph{arXiv preprint arXiv:2205.08084}} (\bibinfo{year}{2022}).
\newblock


\bibitem[Devlin et~al\mbox{.}(2018)]%
        {devlin2018bert}
\bibfield{author}{\bibinfo{person}{Jacob Devlin}, \bibinfo{person}{Ming-Wei Chang}, \bibinfo{person}{Kenton Lee}, {and} \bibinfo{person}{Kristina Toutanova}.} \bibinfo{year}{2018}\natexlab{}.
\newblock \showarticletitle{Bert: Pre-training of deep bidirectional transformers for language understanding}.
\newblock \bibinfo{journal}{\emph{arXiv preprint arXiv:1810.04805}} (\bibinfo{year}{2018}).
\newblock


\bibitem[Dong et~al\mbox{.}(2022)]%
        {dong2022survey}
\bibfield{author}{\bibinfo{person}{Qingxiu Dong}, \bibinfo{person}{Lei Li}, \bibinfo{person}{Damai Dai}, \bibinfo{person}{Ce Zheng}, \bibinfo{person}{Zhiyong Wu}, \bibinfo{person}{Baobao Chang}, \bibinfo{person}{Xu Sun}, \bibinfo{person}{Jingjing Xu}, {and} \bibinfo{person}{Zhifang Sui}.} \bibinfo{year}{2022}\natexlab{}.
\newblock \showarticletitle{A survey for in-context learning}.
\newblock \bibinfo{journal}{\emph{arXiv preprint arXiv:2301.00234}} (\bibinfo{year}{2022}).
\newblock


\bibitem[Edalati et~al\mbox{.}(2022)]%
        {edalati2022krona}
\bibfield{author}{\bibinfo{person}{Ali Edalati}, \bibinfo{person}{Marzieh Tahaei}, \bibinfo{person}{Ivan Kobyzev}, \bibinfo{person}{Vahid~Partovi Nia}, \bibinfo{person}{James~J Clark}, {and} \bibinfo{person}{Mehdi Rezagholizadeh}.} \bibinfo{year}{2022}\natexlab{}.
\newblock \showarticletitle{Krona: Parameter efficient tuning with kronecker adapter}.
\newblock \bibinfo{journal}{\emph{arXiv preprint arXiv:2212.10650}} (\bibinfo{year}{2022}).
\newblock


\bibitem[Fan et~al\mbox{.}(2023)]%
        {fan2023recommender}
\bibfield{author}{\bibinfo{person}{Wenqi Fan}, \bibinfo{person}{Zihuai Zhao}, \bibinfo{person}{Jiatong Li}, \bibinfo{person}{Yunqing Liu}, \bibinfo{person}{Xiaowei Mei}, \bibinfo{person}{Yiqi Wang}, \bibinfo{person}{Jiliang Tang}, {and} \bibinfo{person}{Qing Li}.} \bibinfo{year}{2023}\natexlab{}.
\newblock \showarticletitle{Recommender systems in the era of large language models (llms)}.
\newblock \bibinfo{journal}{\emph{arXiv preprint arXiv:2307.02046}} (\bibinfo{year}{2023}).
\newblock


\bibitem[Gao et~al\mbox{.}(2023)]%
        {gao2023chat}
\bibfield{author}{\bibinfo{person}{Yunfan Gao}, \bibinfo{person}{Tao Sheng}, \bibinfo{person}{Youlin Xiang}, \bibinfo{person}{Yun Xiong}, \bibinfo{person}{Haofen Wang}, {and} \bibinfo{person}{Jiawei Zhang}.} \bibinfo{year}{2023}\natexlab{}.
\newblock \showarticletitle{Chat-rec: Towards interactive and explainable llms-augmented recommender system}.
\newblock \bibinfo{journal}{\emph{arXiv preprint arXiv:2303.14524}} (\bibinfo{year}{2023}).
\newblock


\bibitem[Geng et~al\mbox{.}(2022)]%
        {geng2022recommendation}
\bibfield{author}{\bibinfo{person}{Shijie Geng}, \bibinfo{person}{Shuchang Liu}, \bibinfo{person}{Zuohui Fu}, \bibinfo{person}{Yingqiang Ge}, {and} \bibinfo{person}{Yongfeng Zhang}.} \bibinfo{year}{2022}\natexlab{}.
\newblock \showarticletitle{Recommendation as language processing (rlp): A unified pretrain, personalized prompt \& predict paradigm (p5)}. In \bibinfo{booktitle}{\emph{Proceedings of the 16th ACM Conference on Recommender Systems}}. \bibinfo{pages}{299--315}.
\newblock


\bibitem[Geng et~al\mbox{.}(2023)]%
        {geng2023vip5}
\bibfield{author}{\bibinfo{person}{Shijie Geng}, \bibinfo{person}{Juntao Tan}, \bibinfo{person}{Shuchang Liu}, \bibinfo{person}{Zuohui Fu}, {and} \bibinfo{person}{Yongfeng Zhang}.} \bibinfo{year}{2023}\natexlab{}.
\newblock \showarticletitle{VIP5: Towards Multimodal Foundation Models for Recommendation}.
\newblock \bibinfo{journal}{\emph{arXiv preprint arXiv:2305.14302}} (\bibinfo{year}{2023}).
\newblock


\bibitem[Hasan et~al\mbox{.}(2021)]%
        {hasan2021humor}
\bibfield{author}{\bibinfo{person}{Md~Kamrul Hasan}, \bibinfo{person}{Sangwu Lee}, \bibinfo{person}{Wasifur Rahman}, \bibinfo{person}{Amir Zadeh}, \bibinfo{person}{Rada Mihalcea}, \bibinfo{person}{Louis-Philippe Morency}, {and} \bibinfo{person}{Ehsan Hoque}.} \bibinfo{year}{2021}\natexlab{}.
\newblock \showarticletitle{Humor knowledge enriched transformer for understanding multimodal humor}. In \bibinfo{booktitle}{\emph{Proceedings of the AAAI conference on artificial intelligence}}.
\newblock


\bibitem[Hidasi and Karatzoglou(2017)]%
        {DBLP:journals/corr/HidasiK17}
\bibfield{author}{\bibinfo{person}{Bal{\'{a}}zs Hidasi} {and} \bibinfo{person}{Alexandros Karatzoglou}.} \bibinfo{year}{2017}\natexlab{}.
\newblock \showarticletitle{Recurrent Neural Networks with Top-k Gains for Session-based Recommendations}.
\newblock \bibinfo{journal}{\emph{CoRR}}  \bibinfo{volume}{abs/1706.03847} (\bibinfo{year}{2017}).
\newblock
\showeprint[arXiv]{1706.03847}
\urldef\tempurl%
\url{http://arxiv.org/abs/1706.03847}
\showURL{%
\tempurl}


\bibitem[Hidasi et~al\mbox{.}(2015)]%
        {hidasi2015session}
\bibfield{author}{\bibinfo{person}{Bal{\'a}zs Hidasi}, \bibinfo{person}{Alexandros Karatzoglou}, \bibinfo{person}{Linas Baltrunas}, {and} \bibinfo{person}{Domonkos Tikk}.} \bibinfo{year}{2015}\natexlab{}.
\newblock \showarticletitle{Session-based recommendations with recurrent neural networks}.
\newblock \bibinfo{journal}{\emph{arXiv preprint arXiv:1511.06939}} (\bibinfo{year}{2015}).
\newblock


\bibitem[Hou et~al\mbox{.}(2023a)]%
        {hou2023cold}
\bibfield{author}{\bibinfo{person}{Yupeng Hou}, \bibinfo{person}{Junjie Zhang}, \bibinfo{person}{Zihan Lin}, \bibinfo{person}{Hongyu Lu}, \bibinfo{person}{Ruobing Xie}, \bibinfo{person}{Julian McAuley}, {and} \bibinfo{person}{Wayne~Xin Zhao}.} \bibinfo{year}{2023}\natexlab{a}.
\newblock \bibinfo{title}{Large Language Models are Zero-Shot Rankers for Recommender Systems}.
\newblock
\newblock
\showeprint[arxiv]{2305.08845}~[cs.IR]


\bibitem[Hou et~al\mbox{.}(2023b)]%
        {hou2023large}
\bibfield{author}{\bibinfo{person}{Yupeng Hou}, \bibinfo{person}{Junjie Zhang}, \bibinfo{person}{Zihan Lin}, \bibinfo{person}{Hongyu Lu}, \bibinfo{person}{Ruobing Xie}, \bibinfo{person}{Julian McAuley}, {and} \bibinfo{person}{Wayne~Xin Zhao}.} \bibinfo{year}{2023}\natexlab{b}.
\newblock \showarticletitle{Large language models are zero-shot rankers for recommender systems}.
\newblock \bibinfo{journal}{\emph{arXiv preprint arXiv:2305.08845}} (\bibinfo{year}{2023}).
\newblock


\bibitem[Houlsby et~al\mbox{.}(2019)]%
        {houlsby2019parameter}
\bibfield{author}{\bibinfo{person}{Neil Houlsby}, \bibinfo{person}{Andrei Giurgiu}, \bibinfo{person}{Stanislaw Jastrzebski}, \bibinfo{person}{Bruna Morrone}, \bibinfo{person}{Quentin De~Laroussilhe}, \bibinfo{person}{Andrea Gesmundo}, \bibinfo{person}{Mona Attariyan}, {and} \bibinfo{person}{Sylvain Gelly}.} \bibinfo{year}{2019}\natexlab{}.
\newblock \showarticletitle{Parameter-efficient transfer learning for NLP}. In \bibinfo{booktitle}{\emph{International Conference on Machine Learning}}. PMLR, \bibinfo{pages}{2790--2799}.
\newblock


\bibitem[Hu et~al\mbox{.}(2021)]%
        {hu2021lora}
\bibfield{author}{\bibinfo{person}{Edward~J Hu}, \bibinfo{person}{Yelong Shen}, \bibinfo{person}{Phillip Wallis}, \bibinfo{person}{Zeyuan Allen-Zhu}, \bibinfo{person}{Yuanzhi Li}, \bibinfo{person}{Shean Wang}, \bibinfo{person}{Lu Wang}, {and} \bibinfo{person}{Weizhu Chen}.} \bibinfo{year}{2021}\natexlab{}.
\newblock \showarticletitle{Lora: Low-rank adaptation of large language models}.
\newblock \bibinfo{journal}{\emph{arXiv preprint arXiv:2106.09685}} (\bibinfo{year}{2021}).
\newblock


\bibitem[Hua et~al\mbox{.}(2023)]%
        {hua2023up5}
\bibfield{author}{\bibinfo{person}{Wenyue Hua}, \bibinfo{person}{Yingqiang Ge}, \bibinfo{person}{Shuyuan Xu}, \bibinfo{person}{Jianchao Ji}, {and} \bibinfo{person}{Yongfeng Zhang}.} \bibinfo{year}{2023}\natexlab{}.
\newblock \showarticletitle{UP5: Unbiased Foundation Model for Fairness-aware Recommendation}.
\newblock \bibinfo{journal}{\emph{arXiv preprint arXiv:2305.12090}} (\bibinfo{year}{2023}).
\newblock


\bibitem[Huang et~al\mbox{.}(2023)]%
        {huang2023language}
\bibfield{author}{\bibinfo{person}{Shaohan Huang}, \bibinfo{person}{Li Dong}, \bibinfo{person}{Wenhui Wang}, \bibinfo{person}{Yaru Hao}, \bibinfo{person}{Saksham Singhal}, \bibinfo{person}{Shuming Ma}, \bibinfo{person}{Tengchao Lv}, \bibinfo{person}{Lei Cui}, \bibinfo{person}{Owais~Khan Mohammed}, \bibinfo{person}{Qiang Liu}, {et~al\mbox{.}}} \bibinfo{year}{2023}\natexlab{}.
\newblock \showarticletitle{Language is not all you need: Aligning perception with language models}.
\newblock \bibinfo{journal}{\emph{arXiv preprint arXiv:2302.14045}} (\bibinfo{year}{2023}).
\newblock


\bibitem[Jabeen et~al\mbox{.}(2023)]%
        {jabeen2023review}
\bibfield{author}{\bibinfo{person}{Summaira Jabeen}, \bibinfo{person}{Xi Li}, \bibinfo{person}{Muhammad~Shoib Amin}, \bibinfo{person}{Omar Bourahla}, \bibinfo{person}{Songyuan Li}, {and} \bibinfo{person}{Abdul Jabbar}.} \bibinfo{year}{2023}\natexlab{}.
\newblock \showarticletitle{A review on methods and applications in multimodal deep learning}.
\newblock \bibinfo{journal}{\emph{ACM Transactions on Multimedia Computing, Communications and Applications}} (\bibinfo{year}{2023}).
\newblock


\bibitem[J{\"a}rvelin and Kek{\"a}l{\"a}inen(2002)]%
        {NDCG}
\bibfield{author}{\bibinfo{person}{Kalervo J{\"a}rvelin} {and} \bibinfo{person}{Jaana Kek{\"a}l{\"a}inen}.} \bibinfo{year}{2002}\natexlab{}.
\newblock \showarticletitle{Cumulated gain-based evaluation of IR techniques}.
\newblock \bibinfo{journal}{\emph{ACM Transactions on Information Systems (TOIS)}} \bibinfo{volume}{20}, \bibinfo{number}{4} (\bibinfo{year}{2002}), \bibinfo{pages}{422--446}.
\newblock


\bibitem[Jawahar et~al\mbox{.}(2019)]%
        {jawahar-etal-2019-bert}
\bibfield{author}{\bibinfo{person}{Ganesh Jawahar}, \bibinfo{person}{Beno{\^\i}t Sagot}, {and} \bibinfo{person}{Djam{\'e} Seddah}.} \bibinfo{year}{2019}\natexlab{}.
\newblock \showarticletitle{What Does {BERT} Learn about the Structure of Language?}. In \bibinfo{booktitle}{\emph{Proceedings of the 57th Annual Meeting of the Association for Computational Linguistics}}. \bibinfo{publisher}{Association for Computational Linguistics}, \bibinfo{address}{Florence, Italy}, \bibinfo{pages}{3651--3657}.
\newblock
\urldef\tempurl%
\url{https://doi.org/10.18653/v1/P19-1356}
\showDOI{\tempurl}


\bibitem[Kang and McAuley(2018)]%
        {DBLP:journals/corr/abs-1808-09781}
\bibfield{author}{\bibinfo{person}{Wang{-}Cheng Kang} {and} \bibinfo{person}{Julian~J. McAuley}.} \bibinfo{year}{2018}\natexlab{}.
\newblock \showarticletitle{Self-Attentive Sequential Recommendation}.
\newblock \bibinfo{journal}{\emph{CoRR}}  \bibinfo{volume}{abs/1808.09781} (\bibinfo{year}{2018}).
\newblock
\showeprint[arXiv]{1808.09781}
\urldef\tempurl%
\url{http://arxiv.org/abs/1808.09781}
\showURL{%
\tempurl}


\bibitem[Kaviani and Rahmani(2020)]%
        {kaviani2020emhash}
\bibfield{author}{\bibinfo{person}{Mohadeseh Kaviani} {and} \bibinfo{person}{Hossein Rahmani}.} \bibinfo{year}{2020}\natexlab{}.
\newblock \showarticletitle{Emhash: Hashtag recommendation using neural network based on bert embedding}. In \bibinfo{booktitle}{\emph{2020 6th International Conference on Web Research (ICWR)}}. IEEE, \bibinfo{pages}{113--118}.
\newblock


\bibitem[Lester et~al\mbox{.}(2021)]%
        {lester2021power}
\bibfield{author}{\bibinfo{person}{Brian Lester}, \bibinfo{person}{Rami Al-Rfou}, {and} \bibinfo{person}{Noah Constant}.} \bibinfo{year}{2021}\natexlab{}.
\newblock \showarticletitle{The power of scale for parameter-efficient prompt tuning}.
\newblock \bibinfo{journal}{\emph{arXiv preprint arXiv:2104.08691}} (\bibinfo{year}{2021}).
\newblock


\bibitem[Li et~al\mbox{.}(2023a)]%
        {li2023llava}
\bibfield{author}{\bibinfo{person}{Chunyuan Li}, \bibinfo{person}{Cliff Wong}, \bibinfo{person}{Sheng Zhang}, \bibinfo{person}{Naoto Usuyama}, \bibinfo{person}{Haotian Liu}, \bibinfo{person}{Jianwei Yang}, \bibinfo{person}{Tristan Naumann}, \bibinfo{person}{Hoifung Poon}, {and} \bibinfo{person}{Jianfeng Gao}.} \bibinfo{year}{2023}\natexlab{a}.
\newblock \showarticletitle{Llava-med: Training a large language-and-vision assistant for biomedicine in one day}.
\newblock \bibinfo{journal}{\emph{arXiv preprint arXiv:2306.00890}} (\bibinfo{year}{2023}).
\newblock


\bibitem[Li et~al\mbox{.}(2023b)]%
        {li2023gpt4rec}
\bibfield{author}{\bibinfo{person}{Jinming Li}, \bibinfo{person}{Wentao Zhang}, \bibinfo{person}{Tian Wang}, \bibinfo{person}{Guanglei Xiong}, \bibinfo{person}{Alan Lu}, {and} \bibinfo{person}{Gerard Medioni}.} \bibinfo{year}{2023}\natexlab{b}.
\newblock \showarticletitle{GPT4Rec: A generative framework for personalized recommendation and user interests interpretation}.
\newblock \bibinfo{journal}{\emph{arXiv preprint arXiv:2304.03879}} (\bibinfo{year}{2023}).
\newblock


\bibitem[Li et~al\mbox{.}(2019)]%
        {li2019visualbert}
\bibfield{author}{\bibinfo{person}{Liunian~Harold Li}, \bibinfo{person}{Mark Yatskar}, \bibinfo{person}{Da Yin}, \bibinfo{person}{Cho-Jui Hsieh}, {and} \bibinfo{person}{Kai-Wei Chang}.} \bibinfo{year}{2019}\natexlab{}.
\newblock \showarticletitle{Visualbert: A simple and performant baseline for vision and language}.
\newblock \bibinfo{journal}{\emph{arXiv preprint arXiv:1908.03557}} (\bibinfo{year}{2019}).
\newblock


\bibitem[Li et~al\mbox{.}(2022)]%
        {li2022mlp4rec}
\bibfield{author}{\bibinfo{person}{Muyang Li}, \bibinfo{person}{Xiangyu Zhao}, \bibinfo{person}{Chuan Lyu}, \bibinfo{person}{Minghao Zhao}, \bibinfo{person}{Runze Wu}, {and} \bibinfo{person}{Ruocheng Guo}.} \bibinfo{year}{2022}\natexlab{}.
\newblock \bibinfo{title}{MLP4Rec: A Pure MLP Architecture for Sequential Recommendations}.
\newblock
\newblock
\showeprint[arxiv]{2204.11510}~[cs.IR]


\bibitem[Li et~al\mbox{.}(2021)]%
        {li2021ai}
\bibfield{author}{\bibinfo{person}{Ruilong Li}, \bibinfo{person}{Shan Yang}, \bibinfo{person}{David~A Ross}, {and} \bibinfo{person}{Angjoo Kanazawa}.} \bibinfo{year}{2021}\natexlab{}.
\newblock \showarticletitle{Ai choreographer: Music conditioned 3d dance generation with aist++}. In \bibinfo{booktitle}{\emph{Proceedings of the IEEE/CVF International Conference on Computer Vision}}. \bibinfo{pages}{13401--13412}.
\newblock


\bibitem[Li and Liang(2021)]%
        {li2021prefix}
\bibfield{author}{\bibinfo{person}{Xiang~Lisa Li} {and} \bibinfo{person}{Percy Liang}.} \bibinfo{year}{2021}\natexlab{}.
\newblock \showarticletitle{Prefix-tuning: Optimizing continuous prompts for generation}.
\newblock \bibinfo{journal}{\emph{arXiv preprint arXiv:2101.00190}} (\bibinfo{year}{2021}).
\newblock


\bibitem[Liu et~al\mbox{.}(2021a)]%
        {DBLP:journals/corr/abs-2110-07602}
\bibfield{author}{\bibinfo{person}{Xiao Liu}, \bibinfo{person}{Kaixuan Ji}, \bibinfo{person}{Yicheng Fu}, \bibinfo{person}{Zhengxiao Du}, \bibinfo{person}{Zhilin Yang}, {and} \bibinfo{person}{Jie Tang}.} \bibinfo{year}{2021}\natexlab{a}.
\newblock \showarticletitle{P-Tuning v2: Prompt Tuning Can Be Comparable to Fine-tuning Universally Across Scales and Tasks}.
\newblock \bibinfo{journal}{\emph{CoRR}} (\bibinfo{year}{2021}).
\newblock


\bibitem[Liu et~al\mbox{.}(2021b)]%
        {ptuning}
\bibfield{author}{\bibinfo{person}{Xiao Liu}, \bibinfo{person}{Kaixuan Ji}, \bibinfo{person}{Yicheng Fu}, \bibinfo{person}{Zhengxiao Du}, \bibinfo{person}{Zhilin Yang}, {and} \bibinfo{person}{Jie Tang}.} \bibinfo{year}{2021}\natexlab{b}.
\newblock \showarticletitle{P-Tuning v2: Prompt Tuning Can Be Comparable to Fine-tuning Universally Across Scales and Tasks}.
\newblock \bibinfo{journal}{\emph{CoRR}}  \bibinfo{volume}{abs/2110.07602} (\bibinfo{year}{2021}).
\newblock
\urldef\tempurl%
\url{https://arxiv.org/abs/2110.07602}
\showURL{%
\tempurl}


\bibitem[Liu et~al\mbox{.}(2019)]%
        {liu2019roberta}
\bibfield{author}{\bibinfo{person}{Yinhan Liu}, \bibinfo{person}{Myle Ott}, \bibinfo{person}{Naman Goyal}, \bibinfo{person}{Jingfei Du}, \bibinfo{person}{Mandar Joshi}, \bibinfo{person}{Danqi Chen}, \bibinfo{person}{Omer Levy}, \bibinfo{person}{Mike Lewis}, \bibinfo{person}{Luke Zettlemoyer}, {and} \bibinfo{person}{Veselin Stoyanov}.} \bibinfo{year}{2019}\natexlab{}.
\newblock \showarticletitle{Roberta: A robustly optimized bert pretraining approach}.
\newblock \bibinfo{journal}{\emph{arXiv preprint arXiv:1907.11692}} (\bibinfo{year}{2019}).
\newblock


\bibitem[Lu et~al\mbox{.}(2019)]%
        {lu2019vilbert}
\bibfield{author}{\bibinfo{person}{Jiasen Lu}, \bibinfo{person}{Dhruv Batra}, \bibinfo{person}{Devi Parikh}, {and} \bibinfo{person}{Stefan Lee}.} \bibinfo{year}{2019}\natexlab{}.
\newblock \showarticletitle{Vilbert: Pretraining task-agnostic visiolinguistic representations for vision-and-language tasks}.
\newblock \bibinfo{journal}{\emph{Advances in neural information processing systems}}  \bibinfo{volume}{32} (\bibinfo{year}{2019}).
\newblock


\bibitem[Lu et~al\mbox{.}(2020)]%
        {lu2020twinbert}
\bibfield{author}{\bibinfo{person}{Wenhao Lu}, \bibinfo{person}{Jian Jiao}, {and} \bibinfo{person}{Ruofei Zhang}.} \bibinfo{year}{2020}\natexlab{}.
\newblock \showarticletitle{Twinbert: Distilling knowledge to twin-structured compressed bert models for large-scale retrieval}. In \bibinfo{booktitle}{\emph{Proceedings of the 29th ACM International Conference on Information \& Knowledge Management}}. \bibinfo{pages}{2645--2652}.
\newblock


\bibitem[Lyu et~al\mbox{.}(2023)]%
        {lyu2023macaw}
\bibfield{author}{\bibinfo{person}{Chenyang Lyu}, \bibinfo{person}{Minghao Wu}, \bibinfo{person}{Longyue Wang}, \bibinfo{person}{Xinting Huang}, \bibinfo{person}{Bingshuai Liu}, \bibinfo{person}{Zefeng Du}, \bibinfo{person}{Shuming Shi}, {and} \bibinfo{person}{Zhaopeng Tu}.} \bibinfo{year}{2023}\natexlab{}.
\newblock \showarticletitle{Macaw-LLM: Multi-Modal Language Modeling with Image, Audio, Video, and Text Integration}.
\newblock \bibinfo{journal}{\emph{arXiv preprint arXiv:2306.09093}} (\bibinfo{year}{2023}).
\newblock


\bibitem[Ni et~al\mbox{.}(2019)]%
        {ni-etal-2019-justifying}
\bibfield{author}{\bibinfo{person}{Jianmo Ni}, \bibinfo{person}{Jiacheng Li}, {and} \bibinfo{person}{Julian McAuley}.} \bibinfo{year}{2019}\natexlab{}.
\newblock \showarticletitle{Justifying Recommendations using Distantly-Labeled Reviews and Fine-Grained Aspects}. In \bibinfo{booktitle}{\emph{Proceedings of the 2019 Conference on Empirical Methods in Natural Language Processing and the 9th International Joint Conference on Natural Language Processing (EMNLP-IJCNLP)}}. \bibinfo{publisher}{Association for Computational Linguistics}, \bibinfo{address}{Hong Kong, China}, \bibinfo{pages}{188--197}.
\newblock
\urldef\tempurl%
\url{https://doi.org/10.18653/v1/D19-1018}
\showDOI{\tempurl}


\bibitem[Oord et~al\mbox{.}(2018)]%
        {info}
\bibfield{author}{\bibinfo{person}{Aaron van~den Oord}, \bibinfo{person}{Yazhe Li}, {and} \bibinfo{person}{Oriol Vinyals}.} \bibinfo{year}{2018}\natexlab{}.
\newblock \showarticletitle{Representation learning with contrastive predictive coding}.
\newblock \bibinfo{journal}{\emph{arXiv preprint arXiv:1807.03748}} (\bibinfo{year}{2018}).
\newblock


\bibitem[Peng et~al\mbox{.}(2023)]%
        {peng2023kosmos}
\bibfield{author}{\bibinfo{person}{Zhiliang Peng}, \bibinfo{person}{Wenhui Wang}, \bibinfo{person}{Li Dong}, \bibinfo{person}{Yaru Hao}, \bibinfo{person}{Shaohan Huang}, \bibinfo{person}{Shuming Ma}, {and} \bibinfo{person}{Furu Wei}.} \bibinfo{year}{2023}\natexlab{}.
\newblock \showarticletitle{Kosmos-2: Grounding Multimodal Large Language Models to the World}.
\newblock \bibinfo{journal}{\emph{arXiv preprint arXiv:2306.14824}} (\bibinfo{year}{2023}).
\newblock


\bibitem[Qin and Eisner(2021)]%
        {Qin2021LearningHT}
\bibfield{author}{\bibinfo{person}{Guanghui Qin} {and} \bibinfo{person}{Jas' Eisner}.} \bibinfo{year}{2021}\natexlab{}.
\newblock \showarticletitle{Learning How to Ask: Querying LMs with Mixtures of Soft Prompts}. In \bibinfo{booktitle}{\emph{ACL}}.
\newblock


\bibitem[Qiu et~al\mbox{.}(2021)]%
        {qiu2021u}
\bibfield{author}{\bibinfo{person}{Zhaopeng Qiu}, \bibinfo{person}{Xian Wu}, \bibinfo{person}{Jingyue Gao}, {and} \bibinfo{person}{Wei Fan}.} \bibinfo{year}{2021}\natexlab{}.
\newblock \showarticletitle{U-BERT: Pre-training user representations for improved recommendation}. In \bibinfo{booktitle}{\emph{Proceedings of the AAAI Conference on Artificial Intelligence}}, Vol.~\bibinfo{volume}{35}. \bibinfo{pages}{4320--4327}.
\newblock


\bibitem[Radford et~al\mbox{.}(2021)]%
        {radford2021learning}
\bibfield{author}{\bibinfo{person}{Alec Radford}, \bibinfo{person}{Jong~Wook Kim}, \bibinfo{person}{Chris Hallacy}, \bibinfo{person}{Aditya Ramesh}, \bibinfo{person}{Gabriel Goh}, \bibinfo{person}{Sandhini Agarwal}, \bibinfo{person}{Girish Sastry}, \bibinfo{person}{Amanda Askell}, \bibinfo{person}{Pamela Mishkin}, \bibinfo{person}{Jack Clark}, {et~al\mbox{.}}} \bibinfo{year}{2021}\natexlab{}.
\newblock \showarticletitle{Learning transferable visual models from natural language supervision}. In \bibinfo{booktitle}{\emph{International conference on machine learning}}. PMLR, \bibinfo{pages}{8748--8763}.
\newblock


\bibitem[Radford et~al\mbox{.}(2019)]%
        {radford2019language}
\bibfield{author}{\bibinfo{person}{Alec Radford}, \bibinfo{person}{Jeffrey Wu}, \bibinfo{person}{Rewon Child}, \bibinfo{person}{David Luan}, \bibinfo{person}{Dario Amodei}, \bibinfo{person}{Ilya Sutskever}, {et~al\mbox{.}}} \bibinfo{year}{2019}\natexlab{}.
\newblock \showarticletitle{Language models are unsupervised multitask learners}.
\newblock \bibinfo{journal}{\emph{OpenAI blog}} (\bibinfo{year}{2019}).
\newblock


\bibitem[Raffel et~al\mbox{.}(2019)]%
        {DBLP:journals/corr/abs-1910-10683}
\bibfield{author}{\bibinfo{person}{Colin Raffel}, \bibinfo{person}{Noam Shazeer}, \bibinfo{person}{Adam Roberts}, \bibinfo{person}{Katherine Lee}, \bibinfo{person}{Sharan Narang}, \bibinfo{person}{Michael Matena}, \bibinfo{person}{Yanqi Zhou}, \bibinfo{person}{Wei Li}, {and} \bibinfo{person}{Peter~J. Liu}.} \bibinfo{year}{2019}\natexlab{}.
\newblock \showarticletitle{Exploring the Limits of Transfer Learning with a Unified Text-to-Text Transformer}.
\newblock \bibinfo{journal}{\emph{CoRR}}  \bibinfo{volume}{abs/1910.10683} (\bibinfo{year}{2019}).
\newblock
\showeprint[arXiv]{1910.10683}
\urldef\tempurl%
\url{http://arxiv.org/abs/1910.10683}
\showURL{%
\tempurl}


\bibitem[Rahayu et~al\mbox{.}(2022)]%
        {rahayu2022systematic}
\bibfield{author}{\bibinfo{person}{Nur~W Rahayu}, \bibinfo{person}{Ridi Ferdiana}, {and} \bibinfo{person}{Sri~S Kusumawardani}.} \bibinfo{year}{2022}\natexlab{}.
\newblock \showarticletitle{A systematic review of ontology use in E-Learning recommender system}.
\newblock \bibinfo{journal}{\emph{Computers and Education: Artificial Intelligence}} (\bibinfo{year}{2022}).
\newblock


\bibitem[Rendle et~al\mbox{.}(2012)]%
        {rendle2012bpr}
\bibfield{author}{\bibinfo{person}{Steffen Rendle}, \bibinfo{person}{Christoph Freudenthaler}, \bibinfo{person}{Zeno Gantner}, {and} \bibinfo{person}{Lars Schmidt-Thieme}.} \bibinfo{year}{2012}\natexlab{}.
\newblock \showarticletitle{BPR: Bayesian personalized ranking from implicit feedback}.
\newblock \bibinfo{journal}{\emph{arXiv preprint arXiv:1205.2618}} (\bibinfo{year}{2012}).
\newblock


\bibitem[Resnick and Varian(1997)]%
        {resnick1997recommender}
\bibfield{author}{\bibinfo{person}{Paul Resnick} {and} \bibinfo{person}{Hal~R Varian}.} \bibinfo{year}{1997}\natexlab{}.
\newblock \showarticletitle{Recommender systems}.
\newblock \bibinfo{journal}{\emph{Commun. ACM}} (\bibinfo{year}{1997}).
\newblock


\bibitem[Sanner et~al\mbox{.}(2023)]%
        {sanner2023cold}
\bibfield{author}{\bibinfo{person}{Scott Sanner}, \bibinfo{person}{Krisztian Balog}, \bibinfo{person}{Filip Radlinski}, \bibinfo{person}{Ben Wedin}, {and} \bibinfo{person}{Lucas Dixon}.} \bibinfo{year}{2023}\natexlab{}.
\newblock \bibinfo{title}{Large Language Models are Competitive Near Cold-start Recommenders for Language- and Item-based Preferences}.
\newblock
\newblock
\showeprint[arxiv]{2307.14225}~[cs.IR]


\bibitem[Shin et~al\mbox{.}(2020)]%
        {shin2020autoprompt}
\bibfield{author}{\bibinfo{person}{Taylor Shin}, \bibinfo{person}{Yasaman Razeghi}, \bibinfo{person}{Robert~L Logan~IV}, \bibinfo{person}{Eric Wallace}, {and} \bibinfo{person}{Sameer Singh}.} \bibinfo{year}{2020}\natexlab{}.
\newblock \showarticletitle{Autoprompt: Eliciting knowledge from language models with automatically generated prompts}.
\newblock \bibinfo{journal}{\emph{arXiv preprint arXiv:2010.15980}} (\bibinfo{year}{2020}).
\newblock


\bibitem[Sivapalan et~al\mbox{.}(2014)]%
        {sivapalan2014recommender}
\bibfield{author}{\bibinfo{person}{Sanjeevan Sivapalan}, \bibinfo{person}{Alireza Sadeghian}, \bibinfo{person}{Hossein Rahnama}, {and} \bibinfo{person}{Asad~M Madni}.} \bibinfo{year}{2014}\natexlab{}.
\newblock \showarticletitle{Recommender systems in e-commerce}. In \bibinfo{booktitle}{\emph{2014 World Automation Congress (WAC)}}. IEEE, \bibinfo{pages}{179--184}.
\newblock


\bibitem[Sun et~al\mbox{.}(2019b)]%
        {sun2019videobert}
\bibfield{author}{\bibinfo{person}{Chen Sun}, \bibinfo{person}{Austin Myers}, \bibinfo{person}{Carl Vondrick}, \bibinfo{person}{Kevin Murphy}, {and} \bibinfo{person}{Cordelia Schmid}.} \bibinfo{year}{2019}\natexlab{b}.
\newblock \showarticletitle{Videobert: A joint model for video and language representation learning}. In \bibinfo{booktitle}{\emph{Proceedings of the IEEE/CVF international conference on computer vision}}.
\newblock


\bibitem[Sun et~al\mbox{.}(2019a)]%
        {Sun2019BERT4RecSR}
\bibfield{author}{\bibinfo{person}{Fei Sun}, \bibinfo{person}{Jun Liu}, \bibinfo{person}{Jian Wu}, \bibinfo{person}{Changhua Pei}, \bibinfo{person}{Xiao Lin}, \bibinfo{person}{Wenwu Ou}, {and} \bibinfo{person}{Peng Jiang}.} \bibinfo{year}{2019}\natexlab{a}.
\newblock \showarticletitle{BERT4Rec: Sequential Recommendation with Bidirectional Encoder Representations from Transformer}.
\newblock \bibinfo{journal}{\emph{CIKM}} (\bibinfo{year}{2019}).
\newblock


\bibitem[Tang and Wang(2018)]%
        {caser}
\bibfield{author}{\bibinfo{person}{Jiaxi Tang} {and} \bibinfo{person}{Ke Wang}.} \bibinfo{year}{2018}\natexlab{}.
\newblock \showarticletitle{Personalized top-n sequential recommendation via convolutional sequence embedding}. In \bibinfo{booktitle}{\emph{Proceedings of the eleventh ACM international conference on web search and data mining}}. \bibinfo{pages}{565--573}.
\newblock


\bibitem[Tenney et~al\mbox{.}(2019)]%
        {tenney2019bert}
\bibfield{author}{\bibinfo{person}{Ian Tenney}, \bibinfo{person}{Dipanjan Das}, {and} \bibinfo{person}{Ellie Pavlick}.} \bibinfo{year}{2019}\natexlab{}.
\newblock \showarticletitle{BERT rediscovers the classical NLP pipeline}.
\newblock \bibinfo{journal}{\emph{arXiv preprint arXiv:1905.05950}} (\bibinfo{year}{2019}).
\newblock


\bibitem[Touvron et~al\mbox{.}(2023)]%
        {touvron2023llama}
\bibfield{author}{\bibinfo{person}{Hugo Touvron}, \bibinfo{person}{Thibaut Lavril}, \bibinfo{person}{Gautier Izacard}, \bibinfo{person}{Xavier Martinet}, \bibinfo{person}{Marie-Anne Lachaux}, \bibinfo{person}{Timoth{\'e}e Lacroix}, \bibinfo{person}{Baptiste Rozi{\`e}re}, \bibinfo{person}{Naman Goyal}, \bibinfo{person}{Eric Hambro}, \bibinfo{person}{Faisal Azhar}, {et~al\mbox{.}}} \bibinfo{year}{2023}\natexlab{}.
\newblock \showarticletitle{Llama: Open and efficient foundation language models}.
\newblock \bibinfo{journal}{\emph{arXiv preprint arXiv:2302.13971}} (\bibinfo{year}{2023}).
\newblock


\bibitem[Tsimpoukelli et~al\mbox{.}(2021)]%
        {tsimpoukelli2021multimodal}
\bibfield{author}{\bibinfo{person}{Maria Tsimpoukelli}, \bibinfo{person}{Jacob~L Menick}, \bibinfo{person}{Serkan Cabi}, \bibinfo{person}{SM Eslami}, \bibinfo{person}{Oriol Vinyals}, {and} \bibinfo{person}{Felix Hill}.} \bibinfo{year}{2021}\natexlab{}.
\newblock \showarticletitle{Multimodal few-shot learning with frozen language models}.
\newblock \bibinfo{journal}{\emph{NeurIPS}}  \bibinfo{volume}{34} (\bibinfo{year}{2021}), \bibinfo{pages}{200--212}.
\newblock


\bibitem[Van~der Maaten and Hinton(2008)]%
        {van2008visualizing}
\bibfield{author}{\bibinfo{person}{Laurens Van~der Maaten} {and} \bibinfo{person}{Geoffrey Hinton}.} \bibinfo{year}{2008}\natexlab{}.
\newblock \showarticletitle{Visualizing data using t-SNE.}
\newblock \bibinfo{journal}{\emph{Journal of machine learning research}} (\bibinfo{year}{2008}).
\newblock


\bibitem[Wang and Lim(2023)]%
        {wang2023zero}
\bibfield{author}{\bibinfo{person}{Lei Wang} {and} \bibinfo{person}{Ee-Peng Lim}.} \bibinfo{year}{2023}\natexlab{}.
\newblock \showarticletitle{Zero-Shot Next-Item Recommendation using Large Pretrained Language Models}.
\newblock \bibinfo{journal}{\emph{arXiv preprint arXiv:2304.03153}} (\bibinfo{year}{2023}).
\newblock


\bibitem[Wang et~al\mbox{.}(2019)]%
        {dssm}
\bibfield{author}{\bibinfo{person}{Shuohang Wang}, \bibinfo{person}{Sheng Zhang}, \bibinfo{person}{Yelong Shen}, \bibinfo{person}{Xiaodong Liu}, \bibinfo{person}{Jingjing Liu}, \bibinfo{person}{Jianfeng Gao}, {and} \bibinfo{person}{Jing Jiang}.} \bibinfo{year}{2019}\natexlab{}.
\newblock \bibinfo{title}{Unsupervised Deep Structured Semantic Models for Commonsense Reasoning}.
\newblock
\newblock
\showeprint[arxiv]{1904.01938}~[cs.CL]


\bibitem[Wang et~al\mbox{.}(2023)]%
        {wang2023generative}
\bibfield{author}{\bibinfo{person}{Wenjie Wang}, \bibinfo{person}{Xinyu Lin}, \bibinfo{person}{Fuli Feng}, \bibinfo{person}{Xiangnan He}, {and} \bibinfo{person}{Tat-Seng Chua}.} \bibinfo{year}{2023}\natexlab{}.
\newblock \showarticletitle{Generative recommendation: Towards next-generation recommender paradigm}.
\newblock \bibinfo{journal}{\emph{arXiv preprint arXiv:2304.03516}} (\bibinfo{year}{2023}).
\newblock


\bibitem[Wei et~al\mbox{.}(2022)]%
        {wei2021finetuned}
\bibfield{author}{\bibinfo{person}{Jason Wei}, \bibinfo{person}{Maarten Bosma}, \bibinfo{person}{Vincent~Y Zhao}, \bibinfo{person}{Kelvin Guu}, \bibinfo{person}{Adams~Wei Yu}, \bibinfo{person}{Brian Lester}, \bibinfo{person}{Nan Du}, \bibinfo{person}{Andrew~M Dai}, {and} \bibinfo{person}{Quoc~V Le}.} \bibinfo{year}{2022}\natexlab{}.
\newblock \showarticletitle{Finetuned language models are zero-shot learners}. In \bibinfo{booktitle}{\emph{ICLR}}.
\newblock


\bibitem[Wu et~al\mbox{.}(2021)]%
        {wu2021empowering}
\bibfield{author}{\bibinfo{person}{Chuhan Wu}, \bibinfo{person}{Fangzhao Wu}, \bibinfo{person}{Tao Qi}, {and} \bibinfo{person}{Yongfeng Huang}.} \bibinfo{year}{2021}\natexlab{}.
\newblock \showarticletitle{Empowering news recommendation with pre-trained language models}. In \bibinfo{booktitle}{\emph{Proceedings of the 44th International ACM SIGIR Conference on Research and Development in Information Retrieval}}. \bibinfo{pages}{1652--1656}.
\newblock


\bibitem[Wu et~al\mbox{.}(2023)]%
        {wu2023survey}
\bibfield{author}{\bibinfo{person}{Likang Wu}, \bibinfo{person}{Zhi Zheng}, \bibinfo{person}{Zhaopeng Qiu}, \bibinfo{person}{Hao Wang}, \bibinfo{person}{Hongchao Gu}, \bibinfo{person}{Tingjia Shen}, \bibinfo{person}{Chuan Qin}, \bibinfo{person}{Chen Zhu}, \bibinfo{person}{Hengshu Zhu}, \bibinfo{person}{Qi Liu}, {et~al\mbox{.}}} \bibinfo{year}{2023}\natexlab{}.
\newblock \showarticletitle{A Survey on Large Language Models for Recommendation}.
\newblock \bibinfo{journal}{\emph{arXiv preprint arXiv:2305.19860}} (\bibinfo{year}{2023}).
\newblock


\bibitem[Wu et~al\mbox{.}(2022)]%
        {wu2022graph}
\bibfield{author}{\bibinfo{person}{Shiwen Wu}, \bibinfo{person}{Fei Sun}, \bibinfo{person}{Wentao Zhang}, \bibinfo{person}{Xu Xie}, {and} \bibinfo{person}{Bin Cui}.} \bibinfo{year}{2022}\natexlab{}.
\newblock \showarticletitle{Graph neural networks in recommender systems: a survey}.
\newblock \bibinfo{journal}{\emph{Comput. Surveys}} (\bibinfo{year}{2022}).
\newblock


\bibitem[Xi et~al\mbox{.}(2023)]%
        {xi2023towards}
\bibfield{author}{\bibinfo{person}{Yunjia Xi}, \bibinfo{person}{Weiwen Liu}, \bibinfo{person}{Jianghao Lin}, \bibinfo{person}{Jieming Zhu}, \bibinfo{person}{Bo Chen}, \bibinfo{person}{Ruiming Tang}, \bibinfo{person}{Weinan Zhang}, \bibinfo{person}{Rui Zhang}, {and} \bibinfo{person}{Yong Yu}.} \bibinfo{year}{2023}\natexlab{}.
\newblock \showarticletitle{Towards Open-World Recommendation with Knowledge Augmentation from Large Language Models}.
\newblock \bibinfo{journal}{\emph{arXiv preprint arXiv:2306.10933}} (\bibinfo{year}{2023}).
\newblock


\bibitem[Xu et~al\mbox{.}(2023)]%
        {xu2023multimodal}
\bibfield{author}{\bibinfo{person}{Peng Xu}, \bibinfo{person}{Xiatian Zhu}, {and} \bibinfo{person}{David~A Clifton}.} \bibinfo{year}{2023}\natexlab{}.
\newblock \showarticletitle{Multimodal learning with transformers: A survey}.
\newblock \bibinfo{journal}{\emph{IEEE Transactions on Pattern Analysis and Machine Intelligence}} (\bibinfo{year}{2023}).
\newblock


\bibitem[Zeng et~al\mbox{.}(2023)]%
        {zeng2023matters}
\bibfield{author}{\bibinfo{person}{Yan Zeng}, \bibinfo{person}{Hanbo Zhang}, \bibinfo{person}{Jiani Zheng}, \bibinfo{person}{Jiangnan Xia}, \bibinfo{person}{Guoqiang Wei}, \bibinfo{person}{Yang Wei}, \bibinfo{person}{Yuchen Zhang}, {and} \bibinfo{person}{Tao Kong}.} \bibinfo{year}{2023}\natexlab{}.
\newblock \showarticletitle{What Matters in Training a GPT4-Style Language Model with Multimodal Inputs?}
\newblock \bibinfo{journal}{\emph{arXiv preprint arXiv:2307.02469}} (\bibinfo{year}{2023}).
\newblock


\bibitem[Zhang et~al\mbox{.}(2023b)]%
        {zhang2023video}
\bibfield{author}{\bibinfo{person}{Hang Zhang}, \bibinfo{person}{Xin Li}, {and} \bibinfo{person}{Lidong Bing}.} \bibinfo{year}{2023}\natexlab{b}.
\newblock \showarticletitle{Video-llama: An instruction-tuned audio-visual language model for video understanding}.
\newblock \bibinfo{journal}{\emph{arXiv preprint arXiv:2306.02858}} (\bibinfo{year}{2023}).
\newblock


\bibitem[Zhang et~al\mbox{.}(2021b)]%
        {zhang2021unbert}
\bibfield{author}{\bibinfo{person}{Qi Zhang}, \bibinfo{person}{Jingjie Li}, \bibinfo{person}{Qinglin Jia}, \bibinfo{person}{Chuyuan Wang}, \bibinfo{person}{Jieming Zhu}, \bibinfo{person}{Zhaowei Wang}, {and} \bibinfo{person}{Xiuqiang He}.} \bibinfo{year}{2021}\natexlab{b}.
\newblock \showarticletitle{UNBERT: User-News Matching BERT for News Recommendation.}. In \bibinfo{booktitle}{\emph{IJCAI}}. \bibinfo{pages}{3356--3362}.
\newblock


\bibitem[Zhang et~al\mbox{.}(2023a)]%
        {zhang2023instruction}
\bibfield{author}{\bibinfo{person}{Shengyu Zhang}, \bibinfo{person}{Linfeng Dong}, \bibinfo{person}{Xiaoya Li}, \bibinfo{person}{Sen Zhang}, \bibinfo{person}{Xiaofei Sun}, \bibinfo{person}{Shuhe Wang}, \bibinfo{person}{Jiwei Li}, \bibinfo{person}{Runyi Hu}, \bibinfo{person}{Tianwei Zhang}, \bibinfo{person}{Fei Wu}, {et~al\mbox{.}}} \bibinfo{year}{2023}\natexlab{a}.
\newblock \showarticletitle{Instruction Tuning for Large Language Models: A Survey}.
\newblock \bibinfo{journal}{\emph{arXiv preprint arXiv:2308.10792}} (\bibinfo{year}{2023}).
\newblock


\bibitem[Zhang et~al\mbox{.}(2019)]%
        {zhang2019deep}
\bibfield{author}{\bibinfo{person}{Shuai Zhang}, \bibinfo{person}{Lina Yao}, \bibinfo{person}{Aixin Sun}, {and} \bibinfo{person}{Yi Tay}.} \bibinfo{year}{2019}\natexlab{}.
\newblock \showarticletitle{Deep learning based recommender system: A survey and new perspectives}.
\newblock \bibinfo{journal}{\emph{ACM computing surveys (CSUR)}} (\bibinfo{year}{2019}).
\newblock


\bibitem[Zhang et~al\mbox{.}(2021a)]%
        {Zhang2021}
\bibfield{author}{\bibinfo{person}{Yuhui Zhang}, \bibinfo{person}{Hao Ding}, \bibinfo{person}{Zeren Shui}, \bibinfo{person}{Yifei Ma}, \bibinfo{person}{James Zou}, \bibinfo{person}{Anoop Deoras}, {and} \bibinfo{person}{Hao Wang}.} \bibinfo{year}{2021}\natexlab{a}.
\newblock \showarticletitle{Language models as recommender systems: Evaluations and limitations}. In \bibinfo{booktitle}{\emph{NeurIPS 2021 Workshop on I (Still) Can't Believe It's Not Better}}.
\newblock


\end{thebibliography}

\end{document}